\documentclass[twocolumn]{aastex61}
\usepackage{amsmath}
\usepackage{rotating}
\usepackage[strict]{changepage}
\usepackage{tabularx}

\hypersetup{linkcolor=red,citecolor=blue,filecolor=cyan,urlcolor=magenta}

\submitjournal{ApJ}

\shorttitle{coronal wave interaction with coronal holes}
\shortauthors{Piantschitsch et al.}

\begin{document}

\newcommand{\etal}{{\it et~al.}}
\newcommand{\ie}{{\it i.e.}}
\newcommand{\eg}{{\it e.g.}}
\newcommand{\goes}{{\it GOES}}
\newcommand{\sdo}{{\it SDO}}​

\title{Numerical simulation of coronal waves interacting with coronal holes:\\ I. Basic Features}

\correspondingauthor{Isabell Piantschitsch}
\email{isabell.piantschitsch@uni-graz.at}

\author[0000-0002-0786-7307]{Isabell Piantschitsch}
\affiliation{IGAM/Institute of Physics, University of Graz, Universit\"atsplatz 5, A-8010 Graz, Austria}

\author{Bojan Vr\v{s}nak}
\affiliation{Hvar Observatory, Faculty of Geodesy, Ka\v{c}i\'{c}eva 26, HR-10000 Zagreb, Croatia}

\author{Arnold Hanslmeier}
\affiliation{IGAM/Institute of Physics, University of Graz, Universit\"atsplatz 5, A-8010 Graz, Austria}

\author{Birgit Lemmerer}
\affiliation{IGAM/Institute of Physics, University of Graz, Universit\"atsplatz 5, A-8010 Graz, Austria}

\author{Astrid Veronig}
\affiliation{IGAM/Institute of Physics, University of Graz, Universit\"atsplatz 5, A-8010 Graz, Austria}

\author{Aaron Hernandez-Perez}
\affiliation{IGAM/Institute of Physics, University of Graz, Universit\"atsplatz 5, A-8010 Graz, Austria}

\author{Ja\v{s}a \v{C}alogovi\'{c}}
\affiliation{Hvar Observatory, Faculty of Geodesy, Ka\v{c}i\'{c}eva 26, HR-10000 Zagreb, Croatia}

\author{Tomislav \v{Z}ic}
\affiliation{Faculty of Engineering, University of Rijeka, Vukovarska ul. 58, 51000, Rijeka, Croatia; tzic@riteh.hr}

\begin{abstract}

We developed a new numerical code that is able to perform 2.5D simulations of a magnetohydrodynamic (MHD) wave propagation in the corona, and its interaction with a low density region, such as a coronal hole (CH). We show that the impact of the wave on the CH leads to different effects, such as reflection and transmission of the incoming wave, stationary features at the CH boundary, or formation of a density depletion. We present a comprehensive analysis of the morphology and kinematics of primary and secondary waves, \ie\ we describe in detail the temporal evolution of density, magnetic field, plasma flow velocity, phase speed and position of the wave amplitude. Effects like reflection, refraction and transmisson of the wave strongly support the theory that large scale disturbances in the corona are fast MHD waves and build the major distinction to the competing pseudo-wave theory. The formation of stationary bright fronts was one of the main reasons for the development of pseudo-waves. Here we show that stationary bright fronts can be produced by the interactions of an MHD wave with a CH.  We find secondary waves that are traversing through the CH and we show that one part of these traversing waves leaves the CH again, while another part is being reflected at the CH boundary inside the CH. We observe a density depletion that is moving in the opposite direction of the primary wave propagation. We show that the primary wave pushes the CH boundary to the right, caused by the wave front exerting dynamic pressure on the CH.

\end{abstract}


\section{Introduction} \label{sec:intro}

Coronal waves are large-scale propagating disturbances, which are observable over the entire solar surface and driven either by coronal mass ejections (CME) or alternatively by solar flares (for a comprehensive review see, \eg,\citealt{Vrsnak_Cliver2008}). Due to inconsistencies between the observations and the predicted behaviour of those waves in theory or simulations \citep{Long2017}, two main branches of theories on how to interpret these disturbances emerged. 

On the one hand we have the wave theory, which is based on the idea that these disturbances are MHD waves; in this scenario the disturbances are treated either as slow-mode waves \citep{Wang2009}, soliton waves \citep{Wills-Davey2007} or alternatively as fast-mode waves  \citep[e.g.][]{Wills-Davey_Thompson1999,Wang2000,Wu2001,Ofman2002,Patsourakos2009,Patsourakos_etal.2009,Schmidt_Ofman2010}. Moreover, the theory of fast-mode waves can be split into two interpretations: linear \citep{Thompson1998} and nonlinear wave theory \citep{Vrsnak_Lulic2000,Lulic_etal2013,Warmuth2004,Veronig2010}. In the latter the disturbances are interpreted as large-amplitude pulses. If the amplitude of the wave becomes large enough, the importance of the nonlinear terms increases, \ie\ a steepening of the wave front and a subsequent shock formation is expected. 

On the other hand, competing scenarios to this wave interpretation elaborated on a pseudo-wave theory, which interprets the observed disturbances as a result of the reconfiguration of the coronal magnetic field, due to different physical phenomena, such as Joule heating \citep{Delanee_Hochedez2007}, continuous small-scale reconnection \citep{Attrill2007a,Attrill2007b,van_Driel-Gesztelyi_etal_2008} or stretching of magnetic field lines \citep{Chen_etal2002}. Alternatively, hybrid models try to combine both wave and pseudo-wave interpretations \citep{Chen_etal2002,Chen_etal2005,Zhukov_Auchere2004,Cohen_etal2009,Chen_Wu2011,Downs_etal2011,Cheng_etal2012,Liu_Nitta_etal2010}, where the outer envelope of a CME is interpreted as a pseudo-wave that is followed by a freely propagating fast-mode MHD wave. However, among the different interpretations of these large-scale propagating disturbances, the theory with the most evidence is the fast-mode MHD wave theory \citep{Long2017,Warmuth2015}.

In this paper we focus on the nonlinear wave theory and consider the disturbances as fast-mode MHD waves. In particular, we describe what happens when these waves interact with coronal holes (CHs). In a number of observational cases, authors report such interactions including extreme ultra-violet (EUV) waves moving along the boundaries of CHs \citep{Gopalswamy_etal2009,Kienreich_etal2012}, waves being deflected by active regions \citep{Thompson_etal1999,Delanee_Aulanier1999,Chen_etal2005}, Moreton waves partially penetrating into a CH \citep{Veronig_etal2006}, waves being reflected and refracted at a CH \citep{Kienreich_etal2012,Veronig_etal2008,Long_etal2008,Gopalswamy_etal2009} or even waves being transmitted through a CH \citep{Olmedo2012}. In addition, numerical simulations that are considering the observed disturbances as fast-mode MHD waves \citep{Wang2000,Wu2001,Ofman2002,Downs_etal2011} support these observational facts, since the waves show effects like deflection and refraction when they interact with a CH. 

Effects like reflection, refraction and transmisson of these propagating disturbances strongly support the wave theory and build the major distinction to the competing pseudo-wave theory. Besides other reasons, the existence of stationary bright fronts led to the development of pseudo-wave models. However, recent studies show that fast-mode EUV waves can generate stationary fronts, by passing through a magnetic quasi-separatrix layer \citep{Chen2016}. The present work reveals that stationary features at a CH boundary can be the result of the interaction of a MHD wave with a CH. In our 2D simulations we study in detail the behaviour of a large scale amplitude MHD wave and the subsequent effects after the interaction with a low density region, characterized by high Alfv\'{e}n speed, which are main properties of CHs. We will show that the results of our simulations are consistent with the wave theory.

In Section 2, we introduce the numerical method and describe the setup for the initial conditions. In Section 3 we discuss the morphology of the reflected, traversing and transmitted waves, as well as the time evolution of the stationary features and the density depletion. A detailed analysis of the kinematic measurements of secondary waves and their primary counterparts is presented in Section 4 and we conclude in Section 5.

\section{Numerical Setup} 
\subsection{Algorithm and Equations}

We developed a new numerical MHD code that is
able to perform 2.5D simulations of CH propagation
and its interaction with a low density region, which represents a coronal hole. The code is based on the so called Total Variation Diminishing Lax-Friedrichs (TVDLF) scheme, which was first described and tested by \citet{Toth_Odstrcil1996}. This scheme is a fully explicit method and it is proven to behave well near discontinuities, which is of special importance in our simulations, since we are dealing with discontinuities and shock formation. By applying this TVDLF-method, we achieve second order accuracy in space and time. This second order accuracy ensures that our algorithm is not too diffusive and the use of a so-called slope-limiter guarantees a stable behaviour near discontiniuties and shocks, which usually can hardly be fulfilled by high-order schemes. In order to accomodate the two spatial dimensions in our study, we apply an operator splitting method, which changes the order of dimension in each time step and thus ensures that our scheme stays second order \citep{Toro1999}. The resolution in our simulation is $500\times300$ grid points and
the dimensionless length of the computational box is equal to 1.0 both
in $x$- and $y$-direction, \ie\ $\triangle x=0.002$, $\triangle y\approx0.003$.
The wave is propagating in positive $x$-direction and we use
transmissive boundary conditions at the right and left boundary of
the computational box. The algorithm was basically developed as a
two-fluid code, taking into account a neutral fluid, an ionized fluid
and their interactions such as collisional or recombination effects.
For our purposes it is sufficient to apply the single-fluid MHD part
of the code, and therefore neglect the equations for the neutral fluid.

The following set of equations with standard notations for the variables
describes the two-dimensional MHD model we use in flux conservative
form.

\textcolor{black}{
\begin{equation}
\frac{\partial{\color{black}\rho}}{\partial t}+\nabla\cdot(\rho v)=0
\end{equation}
}

\textcolor{black}{
\begin{equation}
\frac{\partial(\rho v)}{\partial t}+\nabla\cdot\left(\rho vv\right)-J\times B+\nabla p=0
\end{equation}
}

\textcolor{black}{
\begin{equation}
\frac{\partial B}{\partial t}-\nabla\times(v\times B)=0
\end{equation}
}

\textcolor{black}{
\begin{equation}
\frac{\partial e}{\partial t}+\nabla\cdot\left[\left(e+p\right)v\right]=0
\end{equation}
}where the plasma energy $e$ is given by 
\textcolor{black}{
\begin{equation}
e=\frac{p}{\gamma-1}+\frac{\rho|v|^{2}}{2}+\frac{|B|^{2}}{2}
\end{equation}
}and $\gamma=5/3$ denotes the adiabatic index.

\subsection{Initial Conditions}

We consider an idealized case where we assume to have a homogeneous
magnetic field in vertical direction and zero pressure all over the
computational box, \ie\ $B_{x}=B_{y}=0$, $p=0$ for
$0\leq x,y\leq1$. The initial setup for all variables in our simulation
looks as follows:

\begin{equation}
     \rho(x) = 
    \begin{cases}
        \Delta\rho\cdot cos^2(\pi\frac{x-x_0}{\Delta x})+\rho_0 & 0.05\leq x\leq0.15 \\
        \qquad \qquad 0.1 & \:\:0.4\leq x\leq0.6 \\
        \qquad \qquad1.0 & \:\:\qquad\text{else}
    \end{cases}
\end{equation}

\begin{equation}
    v_x(x) = 
    \begin{cases}
        2\cdot \sqrt{\frac{\rho(x)}{\rho_0}} -2.0& 0.05\leq x\leq0.15 \\
        \:\:0 & \:\:\qquad\text{else}
    \end{cases}
\end{equation}

\begin{equation}
    B_z(x) = 
    \begin{cases}
        \:\:\rho(x) & 0.05\leq x\leq0.15 \\
        \:\: 1.0 & \:\:\qquad\text{else}
    \end{cases}
\end{equation}

\begin{equation}
B_{x}=B_{y}=0,\qquad0\leq x\leq1
\end{equation}

\begin{equation}
v_{y}=v_{z}=0,\qquad0\leq x\leq1
\end{equation}

where $\rho_{0}=1.0$, $\triangle\rho=0.5$,
$x_{0}=0.1$, $\triangle x=0.1$.

\begin{figure}[ht!]
\centering\includegraphics[width=0.45\textwidth]{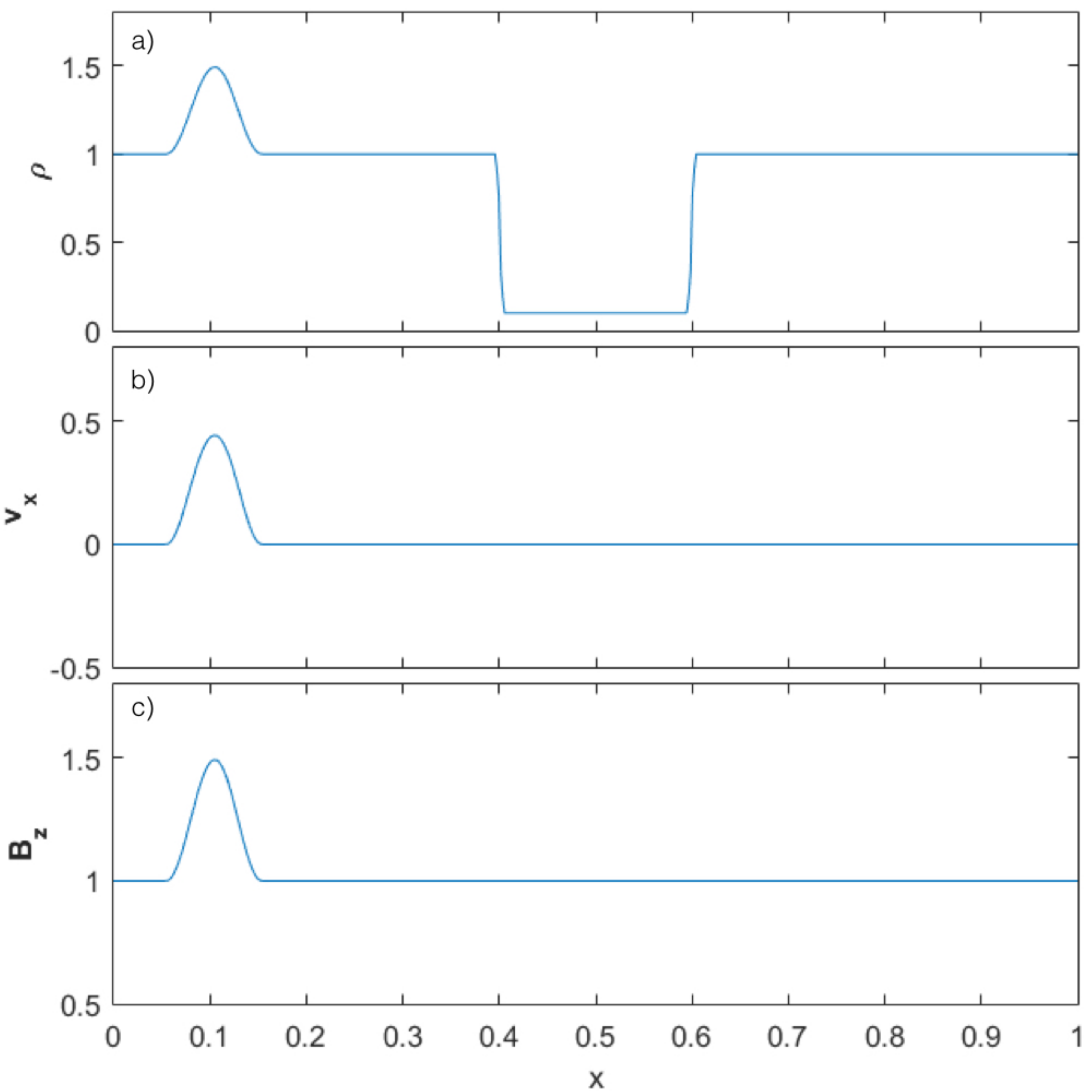}
\caption{Vertical cut through the two-dimensional initial conditions at $y=0.5$ for density $\rho$, plasma flow velocity $v_x$ and magnetic field in $z$-direction $B_z$.}
\label{InitCond_1D}
\end{figure}

\begin{figure*}[ht!]
\centering \includegraphics[width=\textwidth]{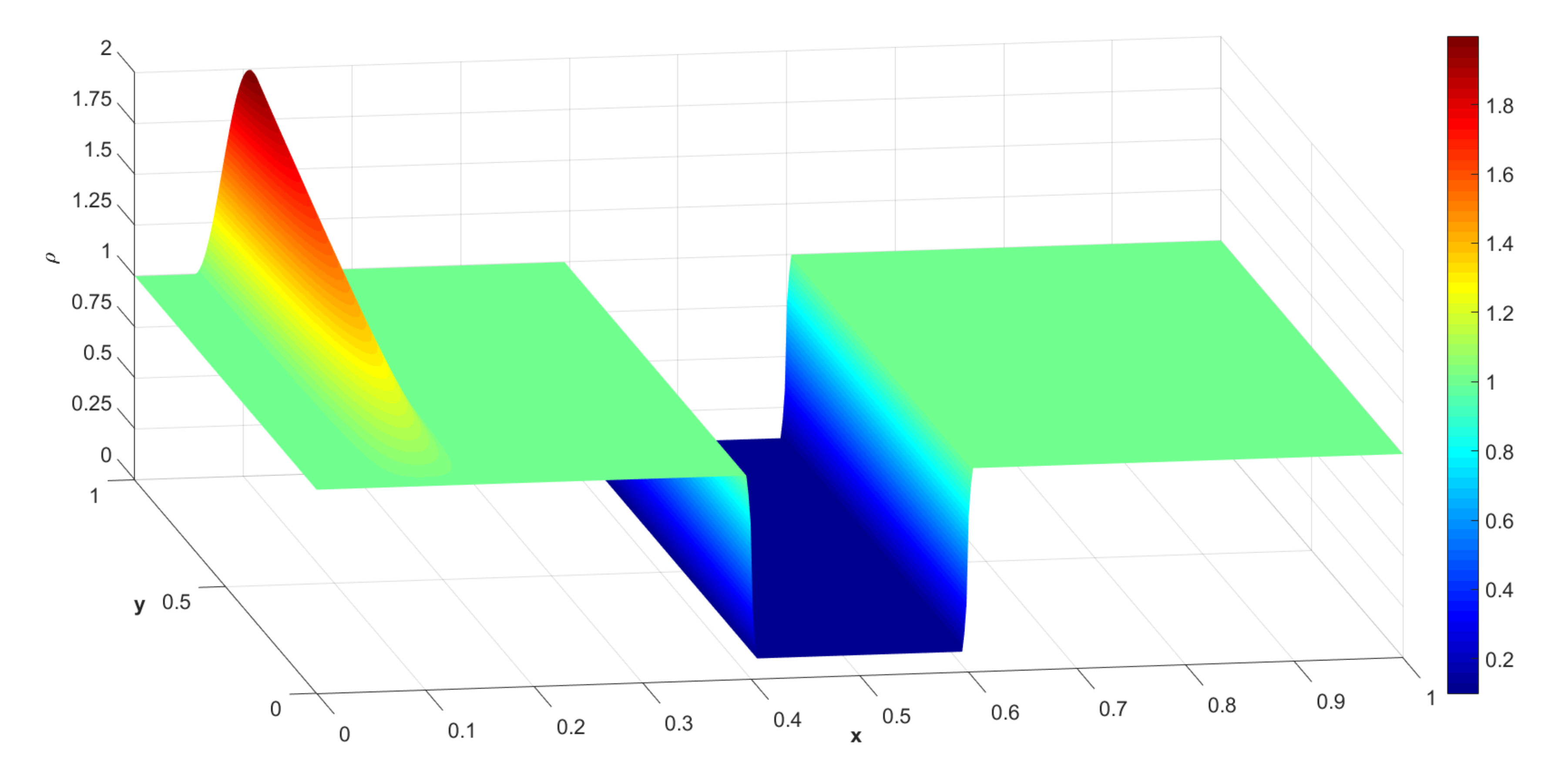}
\caption{Initial two-dimensional density distribution, showing a linearly increasing amplitude in positive $y$-direction from $\rho=1.0$ to $\rho=2.0$ and a coronal hole density of $\rho_{CH}=0.1$. }
\label{initcond_2D}
\end{figure*}

Figure \ref{InitCond_1D} shows a vertical cut through the 2D initial conditions at the
position $y=0.5$ for density $\rho$, $z$-component of the magnetic field $B_{z}$ and plasma flow velocity in $x$-direction $v_{x}$. In Figure \ref{InitCond_1D}a we can see a density drop from $\rho=1.0$ to $\rho=0.1$ in the range $0.4\leq x\leq0.6$, which represents the coronal hole in our simulations. Between the positions $x=0.05$ and $x=0.15$ we created a wave with an initial amplitude of $\rho=1.5$ at $y=0.5$ (detailed description in formulas (6)-(10)). The background density is equal to $1.0$ for the whole computational grid. Figure \ref{InitCond_1D}b and \ref{InitCond_1D}c show how the initial $B_{z}$ and $v_{x}$ are defined as a function of $\rho$ in the range $0.05\leq x\leq0.15$ (see formulas (6) - (8)). The background magnetic field is homogeneous in $z$-direction, \ie\ $B_{z,B}=1.0$, whereas $B_{x}=B_{y}=0$ all over the computational box. The plasma velocity in $x$-direction is equal to zero outside the range where the wave is created and also equal to zero for the $y$- and $z$-velocity components over the whole grid.
Figure \ref{initcond_2D} shows the 2D initial conditions for the density. It can be seen how the amplitude of the density increases linearly from a value of $\rho=1.0$ at $y=0$ to $\rho=2.0$ at $y=1$ in positive $y$-direction. This initial setup enables us to study the wave propagation for different amplitudes of the incoming wave by running the code only once. In our study the CH is constructed as a border along the
$y$-direction of width $0.2$ in the range $0.4\leq x\leq0.6$.

\section{Morphology}

\begin{figure*}[ht]
\centering \includegraphics[width=\textwidth]{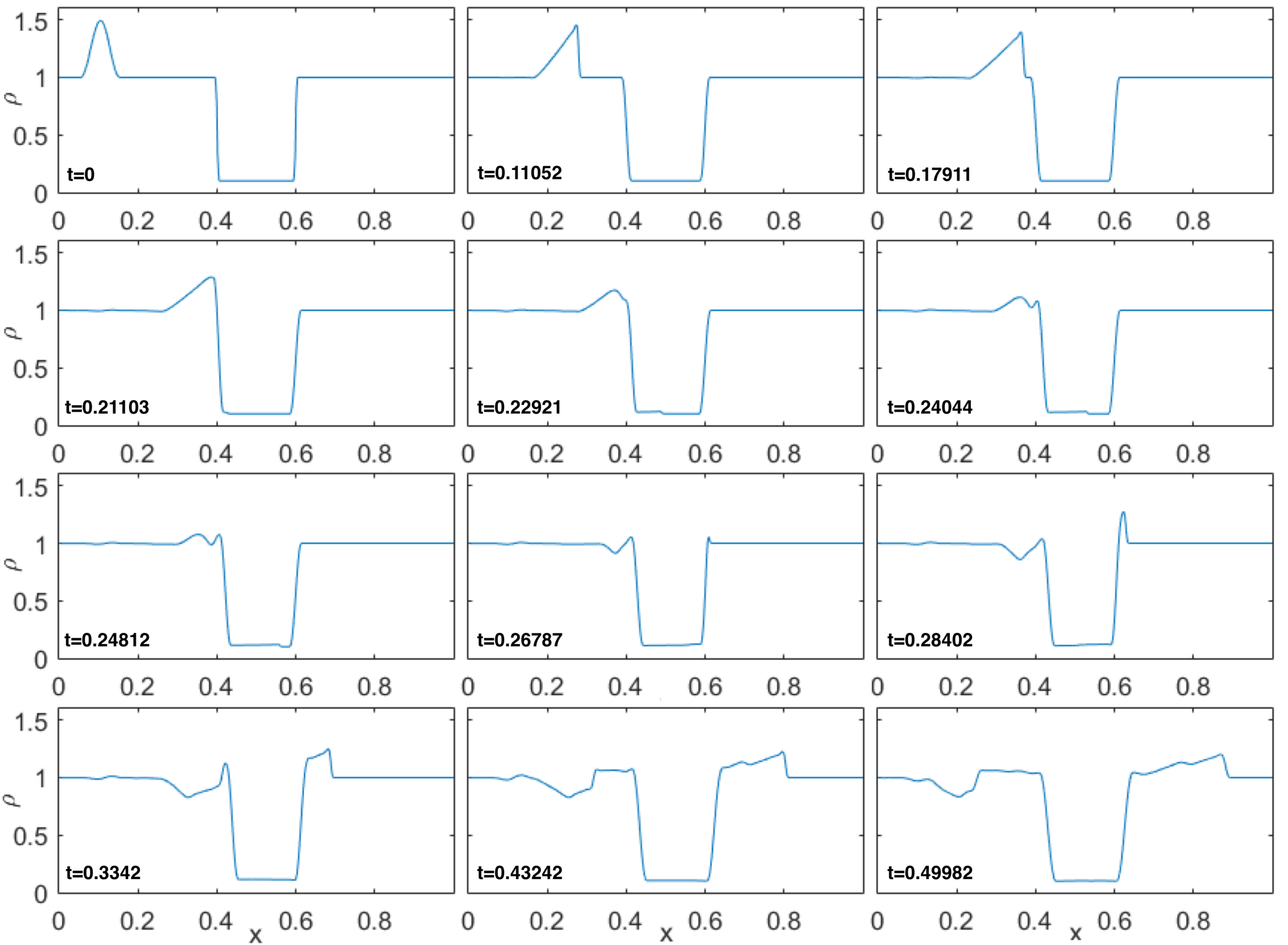}  
\caption{Temporal evolution of the density distribution, starting at the beginning of the run at $t=0$ and ending at  $t=0.5$.}
\label{morphology_2}
\end{figure*}

\begin{figure*}[ht]
\centering \includegraphics[width=\textwidth]{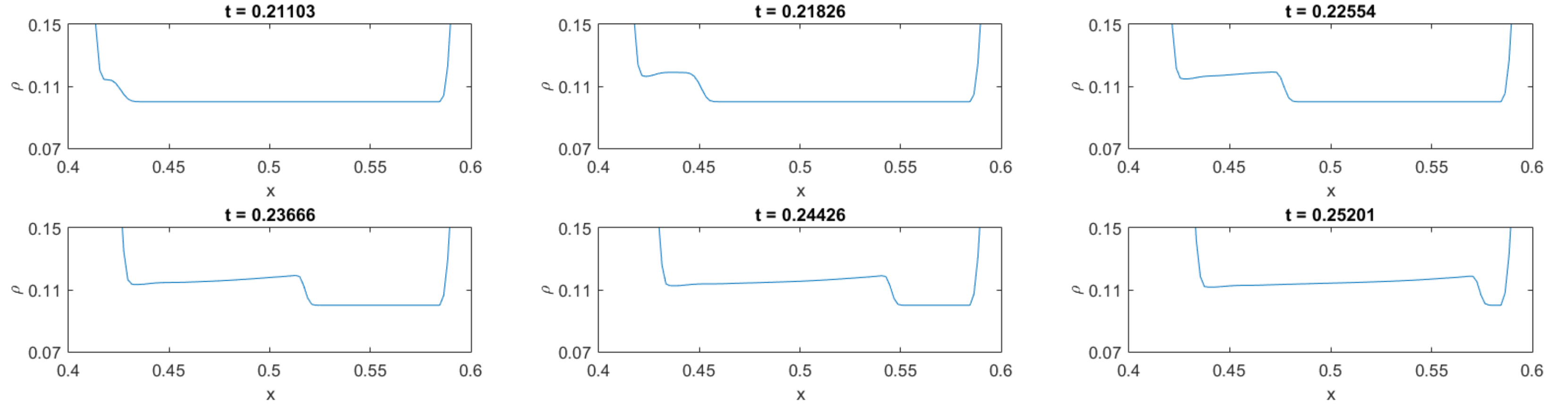}  
\caption{Temporal evolution of the density distribution of the first traversing wave inside the CH, starting shortly after the primary wave has entered the CH ($t=0.21103$) and ending before one part of the wave gets reflected inside the CH and another part of the wave leaves the CH ($t=0.25201$).}
\label{morphology_IN_CH}
\end{figure*}

\begin{figure*}[ht]
\centering \includegraphics[width=\textwidth]{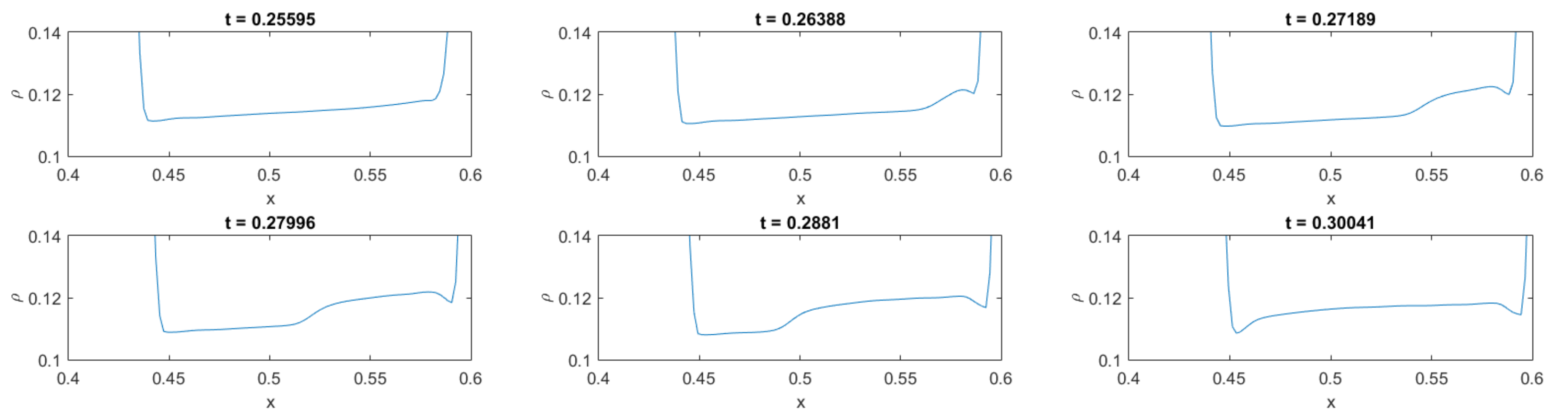}  
\caption{Temporal evolution of the density distribution of the second traversing wave which appears as a result of the reflection of the first traversing wave at the right CH boundary inside the CH and is moving in negative $x$-direction. }
\label{morphology_IN_CH_no2}
\end{figure*}

\begin{figure*}[ht]
\centering \includegraphics[width=\textwidth]{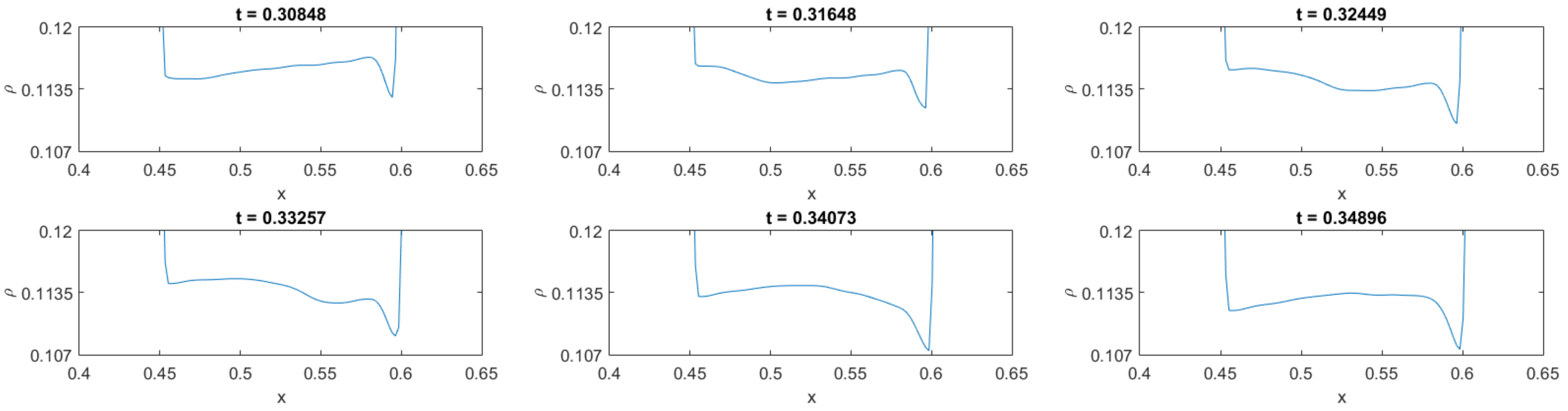}  
\caption{Temporal evolution of the density distribution of the third traversing wave which is a consequence of the reflection of the second traversing at the left CH boundary inside the CH. }
\label{morphology_IN_CH_no3}
\end{figure*}

\begin{sidewaysfigure*}[ht!]
\begin{center}
\vspace*{9cm}
\hspace*{-2cm}
\includegraphics[width=24.5cm,height=17cm]{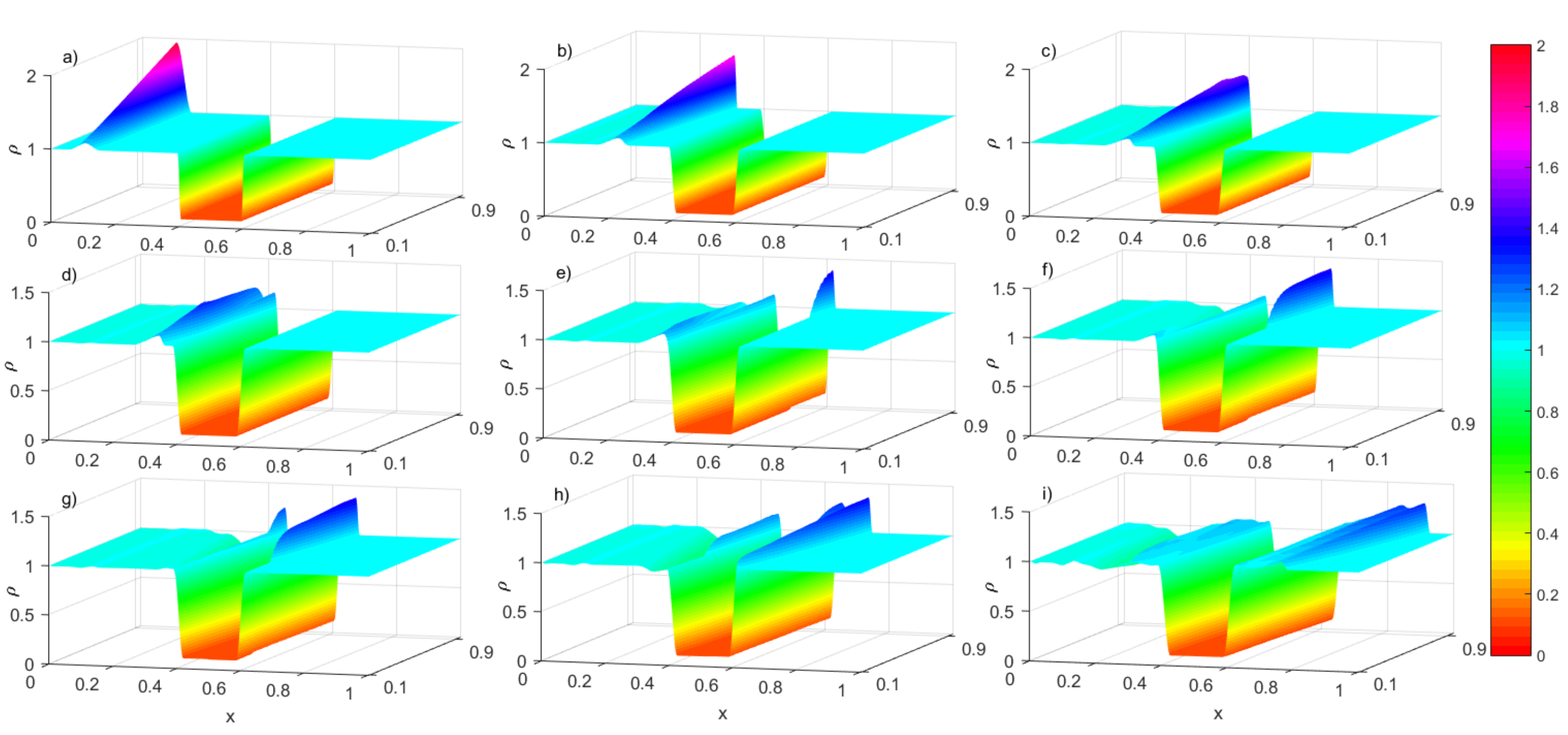}
\end{center}
\hspace*{-2cm}
\caption{Temporal evolution of the two-dimensional density distribution for different initial amplitude values. a) Initial density configuration for the propagating wave and the CH. b--h) Wave propagation before, during and after entering the CH. i) Final density distribution at the end of the simulation run at $t=0.5$.}
\label{morphology_1}
\end{sidewaysfigure*}

In Figure \ref{morphology_2} we have plotted a vertical cut through the $xz$-plane of our simulations at $y=0.5$, showing the density distribution at twelve different time steps. We can see the temporal evolution
of the incoming wave (hereafter primary wave), its interaction with a CH and the subsequently evolving secondary waves (\ie\ reflected, transmitted and traversing waves). We find stationary effects at the left CH boundary and a density depletion at the left side of the CH, moving in negative $x$-direction. Furthermore, we observe that the primary wave pushes the CH boundary in the direction of wave propagation, due to the wave front exerting dynamic pressure on the CH.

\subsection{Primary Wave} 
The primary wave, having an initial density amplitude of $\rho=1.5$ (seen at $t=0$ in Figure \ref{morphology_2}), is moving in positive $x$-direction toward the CH boundary ($t=0.11052$ and $t=0.17911$). As the wave moves towards the CH, a decrease in its amplitude and a broadening of its width occurs. At $t=0.11052$ one can already observe a steepening of the wave that is subsequently developing into a shock ($t=0.17911$).

\subsection{Secondary Waves}

Due to the reduced density in the CH, the amplitude quickly decreases
after the primary wave has started entering the CH (seen at $t=0.21103$ in Figure \ref{morphology_2}). At $t=0.22921$ we observe an immediate first reflection of the primary wave at the left CH boundary (at $x\approx0.4$), which remains observable as a stationary feature at $t=0.24044$ and $t=0.24812$. During this time interval, segments in the rear of the primary wave keep entering the CH. Due to this superposition of the flow associated with the primary wave and the reflection, this peak remains at the same position. After the primary wave has fully entered the CH, the first reflection, which is followed by a density depletion, is able to propagate in negative $x$-direction but with a very low amplitude that can hardly be distinguished from the background density (see $t=0.3342$, $t=0.43242$ and $t=0.49982$ in Figure \ref{morphology_2}, for a detailed description of density depletion see Section 3.4). How the traversing wave with low amplitude is propagating through the CH, is evident at $t=0.22921$, $t=0.24044$ and $t=0.24812$ in Figure \ref{morphology_2}. At $t=0.26787$  the traversing wave is leaving the CH again, when the amplitude of the wave increases up to a value of about $\rho=1.3$ and the transmitted wave keeps moving toward the positive $x$-direction until the end of the simulation run (seen at $t=0.28402$, $t=0.3342$, $t=0.43242$ and $t=0.49982$ in Figure \ref{morphology_2}). Due to the large density drop at $0.4\leq x\leq0.6$, the amplitude of the traversing wave can hardly be seen in Figure \ref{morphology_2}. In Figure \ref{morphology_IN_CH}, Figure \ref{morphology_IN_CH_no2} and Figure \ref{morphology_IN_CH_no3} we zoom in on the region $0.4\leq x\leq0.6$ of the density distribution from $t=0.21103$ to $t=0.34896$, which is the time interval where the wave is traversing back and forth within the CH. Figure \ref{morphology_IN_CH} shows that the wave is moving with an approximately constant amplitude of $\rho\approx0.11$ through the CH. One part of this traversing wave is being reflected at the right CH boundary and continues propagating in negative $x$-direction inside the CH (see Figure \ref{morphology_IN_CH_no2}). The other part leaves the CH again and keeps moving in positive $x$-direction (seen at $t=0.28402$, $t=0.3342$, $t=0.43242$ and $t=0.49982$ Figure \ref{morphology_2}). Shortly after this second (reflected) traversing wave has reached the left CH boundary we observe a stationary feature at the CH boundary (seen at $t=0.3342$ and $x\approx0.4$ in Figure \ref{morphology_2}, for a detailed description see Section 3.3) and, subsequently, a reflection moving in negative $x$-direction, at an approximate constant amplitude of $\rho\approx1.07$ (seen at $t=0.43242$ and $t=0.49982$ in Figure \ref{morphology_2}). This second reflection is a result of the traversing wave partially exiting the CH at the left boundary. Again, one part of the traversing wave is being reflected at the CH boundary and starts moving toward positive $x$-direction inside the CH (see Figure \ref{morphology_IN_CH_no3}). After reaching the right CH boundary, the traversing wave leaves the CH and results in a subwave of the first transmitted wave, seen as a second moving peak within the transmitted wave (see  Figure \ref{morphology_2} at $t=0.43242$ and $t=0.49982$). We can also see in Figure \ref{morphology_IN_CH} and Figure \ref{morphology_IN_CH_no2} how the CH boundary is pushed into the propagation direction of the incoming wave. We can see that this shift of the left CH boundary stops at $t\approx0.3$, which is approximately the time at which the whole primary wave has completed the entry phase into the CH.

\subsection{Stationary Features}
At $t=0.22921$ in Figure \ref{morphology_2} we start observing a first stationary feature which appears as a stationary peak at the left hand side of the CH. This peak is caused by the reflection of the primary wave on the CH boundary and lasts only for a short time. It can be seen as an immediate increase of density at $x\approx0.4$ (see Figure \ref{morphology_2} at $t=0.24044$ and $t=0.24812$). This first stationary feature at the left CH boundary occurs while the amplitude of the primary wave is still decreasing and segments in the rear of the primary wave are still entering the CH. At $t=0.3342$ and $t=0.43242$ in Figure \ref{morphology_2} we observe a second stationary feature at the left CH boundary ($t=0.3342$ and $t=0.4324$) which occurs immediately after the traversing wave inside the CH has reached the left CH boundary again. It also appears as a stationary peak but at $x\approx0.43$, and lasts longer than the aforementioned first stationary feature. It remains observable while the second reflection is moving in negative $x$-direction (seen at $t=0.43242$).

\subsection{Density Depletion}

At $t=0.24044$ in Figure \ref{morphology_2} one can observe the beginning of the evolution of a density depletion at $x\approx0.39$, located at the left side of the first stationary feature. The minimum value of the density depletion decreases, finally attaining a value smaller than $\rho=1.0$ ($t=0.26787$) and propagating into negative $x$-direction, ahead of the second reflection (seen at $t=0.3342$, $t=0.43242$ and $t=0.49982$ in Figure \ref{morphology_2}).

\subsection{2D Morphology}

Figure \ref{morphology_1} shows the 2D temporal evolution of the density distribution including all cases with different initial amplitudes from $\rho=1.0$ at $y=0$ to $\rho=2.0$ at $y=1$. In Figure \ref{morphology_1}a we see the initial 2D setup at the beginning of the simulation run. The time evolution of the primary wave is shown in Figures \ref{morphology_1}a, \ref{morphology_1}b and \ref{morphology_1}c, where the waves with larger initial amplitude move faster toward the CH boundary than those with smaller initial amplitude. In Figure \ref{morphology_1}c we see that the waves with the largest initial amplitudes have already reached the CH boundary, while the waves with smaller initial amplitudes are still moving in the direction of the CH. Figure \ref{morphology_1}d already shows the appearance of the first stationary feature at the left CH boundary for larger amplitude waves whereas the smaller amplitude waves are still moving toward the CH. Moreover, we can observe in Figure \ref{morphology_1}e that the larger amplitude waves are already leaving the CH again while the lower amplitude waves are still entering the CH. In Figure \ref{morphology_1}g we see on the one hand that almost all waves have left the CH again and propagate further toward the positive $x$-direction. On the other hand we observe the appearance of the second stationary feature at the left CH boundary for the larger amplitude waves. In Figure \ref{morphology_1}h and \ref{morphology_1}i we observe the transmitted waves propagating in positive $x$-direction and the density depletion and second reflection moving in negative $x$-direction.

\section{Kinematics}

\subsection{Simulations without CH}

\begin{figure*}[ht!]
\centering \includegraphics[width=\textwidth, height=6cm]{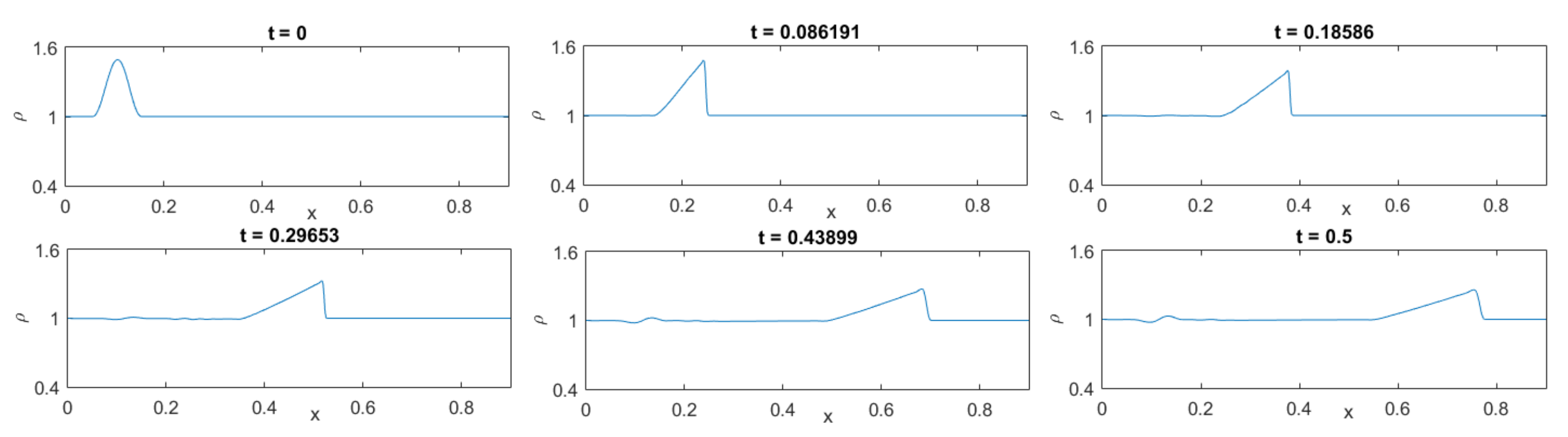}  
\caption{Density distribution of the wave propagation without any interaction with a CH at six different timesteps.}
\label{morph_no_CH}
\end{figure*}

At the beginning of our kinematic analysis we consider a case without
a region of lower density in the initial simulation setup (seen at $t=0$ in Figure \ref{morph_no_CH}), that is, we will describe the behaviour of the wave propagation in positive $x$-direction without any interaction
with an obstacle like a CH, in order to have a reference behaviour of the simple wave propagation. Since numeric simulations can lead, depending on the algorithm, to numerical diffusion and other numerical effects, we will first pay attention to this simple case, in order to be able to interpret reliably the situation including a CH. Figure \ref{morph_no_CH} shows density plots at six different time steps, that
describe the temporal evolution of the wave without any interaction
with a CH. At $t=0$ we see the initial setup of the density distribution with a starting amplitude of $\rho=1.5$. From $t=0$ until the end of the run at $t=5$ we see that the amplitude decreases from the initial value of $\rho=1.5$ down to a value of $\rho\approx1.25$, while at the same time the width of the wave is getting larger, starting at $width_{wave}=0.1$ ($t=0$) and reaching a value of $width_{wave}=0.2$ ($t=0.5$). At $t=0.086191$ we start observing a steepening of the wave
that is followed by a shock formation. The oscillations that can be seen in Figure \ref{morph_no_CH} at $x\approx0.1$ from $t=0.18586$ until the end of the run occur most probably due to numerical effects.

\begin{figure}[ht!]
\centering \includegraphics[width=0.5\textwidth]{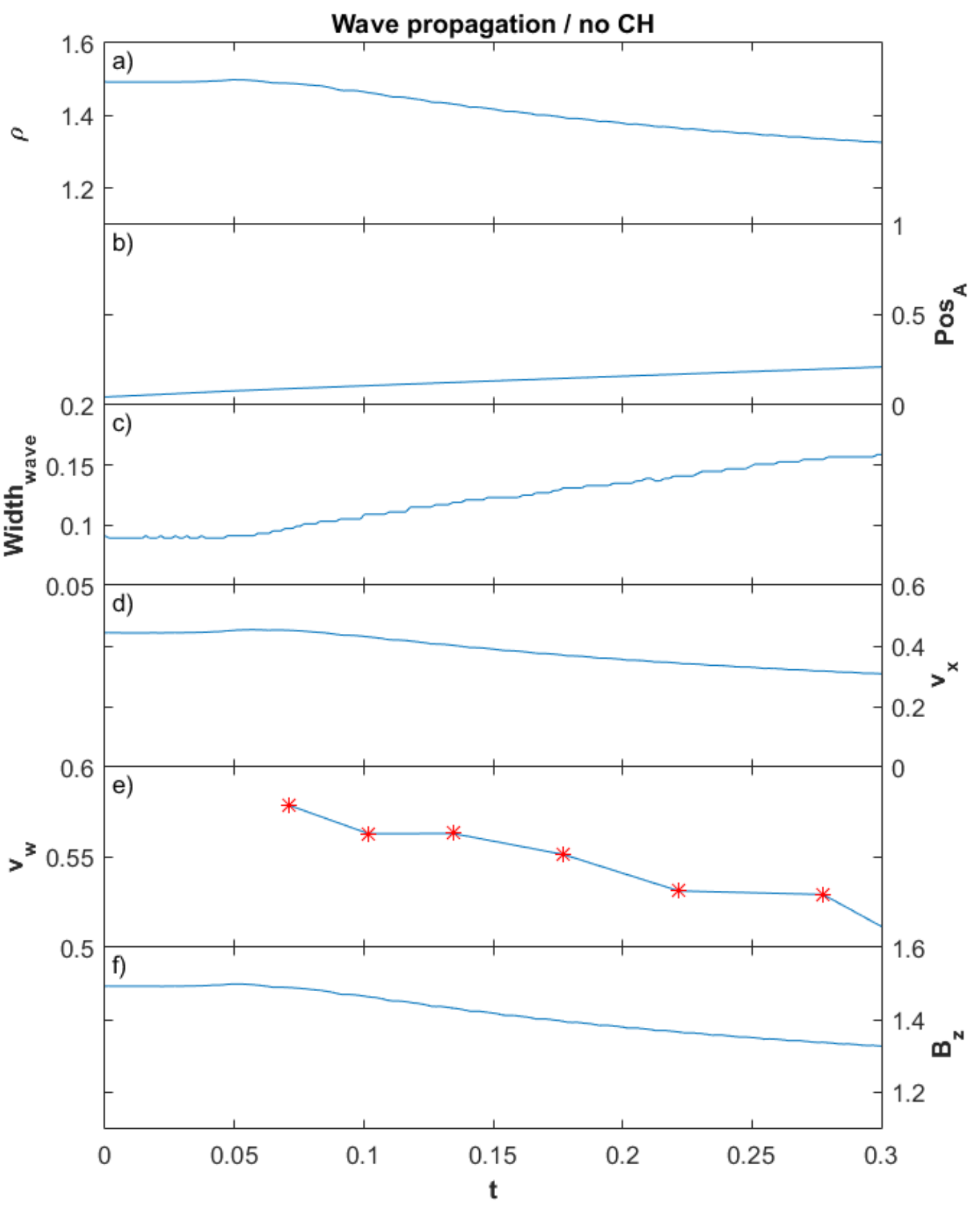}
\caption{From top to bottom: Density, position of the amplitude, width of the wave, plasma flow velocity, phase velocity and magnetic field of the wave propagating in positive $x$-direction without interaction with a CH.}
\label{kinem_no_CH}
\end{figure}

In Figure \ref{kinem_no_CH} we present the evolution of the
following variables: density $\rho$, position of the amplitude $Pos_{A}$, width of the wave $width_{wave}$, 
plasma flow velocity $v_{x}$, actual phase velocity of the wave crest $v_{w}$ and the magnetic
field component $B_{z}$. Figures \ref{kinem_no_CH}a, \ref{kinem_no_CH}d and \ref{kinem_no_CH}f show that the amplitude of density, plasma flow velocity and magnetic field component in $z$-direction stay approximately constant at values $\rho\approx1.5$, $v_{x}\approx0.4$ and $B_{z}\approx1.5$ until $t\approx0.05$. The values then start decreasing and finally reach values of $\rho\approx1.35$, $v_{x}\approx0.3$ and $B_{z}\approx1.35$ at $t=0.3$. At the same time when $\rho$, $v_{x}$ and $B_{z}$ start decreasing, the width of the wave is getting larger, starting at $width_{wave}\approx0.1$ and reaching a value of $width_{wave}\approx0.16$ at $t=0.3$. We observe in Figure \ref{kinem_no_CH}e a decrease of the phase speed from $v_{w}\approx1.3$ at $t\approx0.07$ to $v_{w}\approx0.7$ at $t=0.3$, similar to the behaviour observed in Figures \ref{kinem_no_CH}a, \ref{kinem_no_CH}d and \ref{kinem_no_CH}f. The plot of the amplitude position in Figure \ref{kinem_no_CH}b shows how the wave is propagating forward in positive $x$-direction. Furthermore, in Figure \ref{kinem_no_CH}e we find a decrease of the phase speed from about $v_{w}=0.45$ (at $t=0$) to $v_{w}=0.25$ (at $t=0.5$).

\subsection{Simulations including a CH}

\subsubsection{Primary Wave}

\begin{figure}[ht!]
\centering \includegraphics[width=0.5\textwidth]{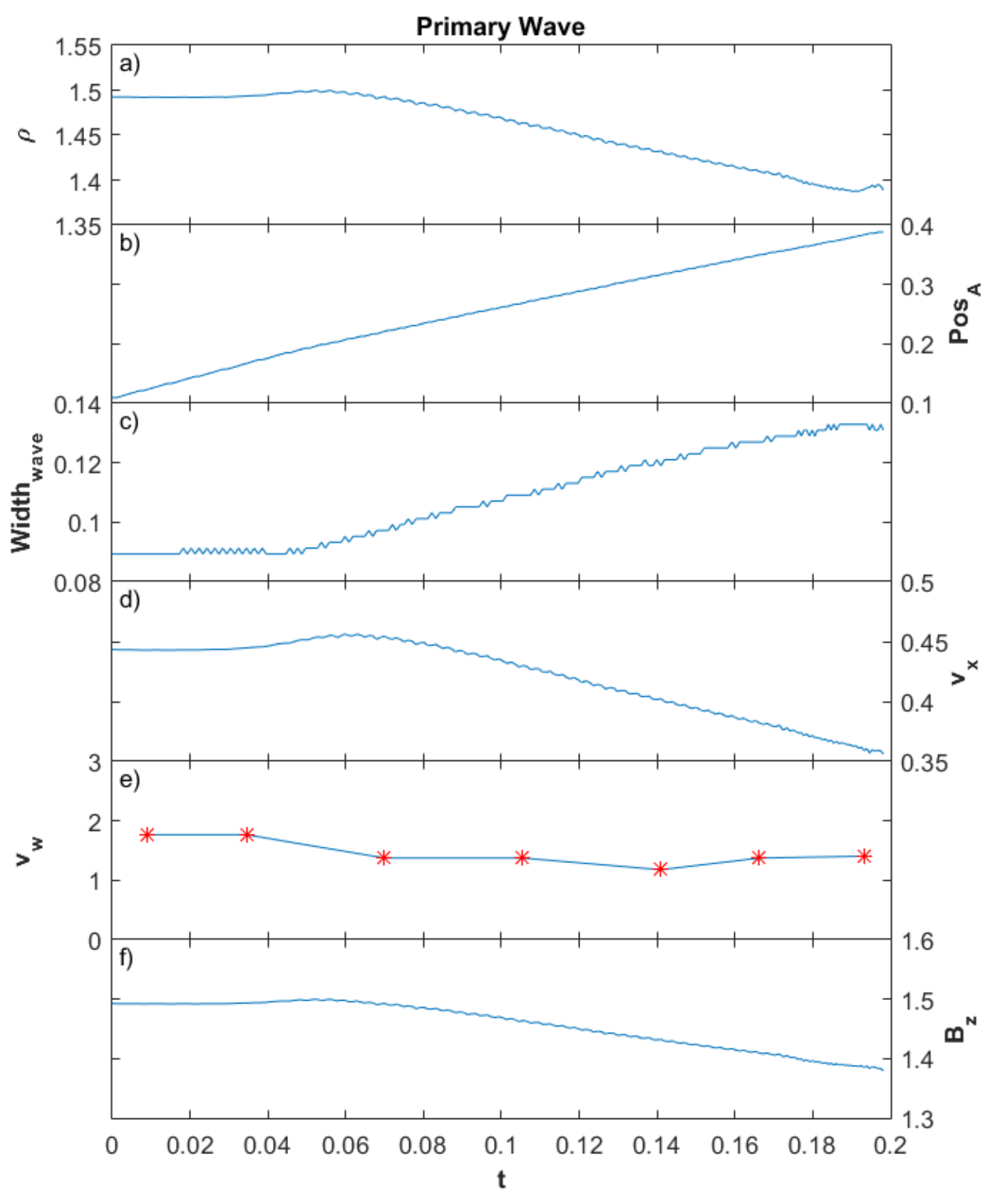}
\caption{From top to bottom: Density, position of the amplitude, width of the wave, plasma flow velocity, phase velocity and magnetic field of the primary wave, from the beginning of the run ($t=0$) until the time when the wave is entering the CH ($t=0.2$). }
\label{kin_prim_wave}
\end{figure}

The temporal evolution of the peak values of density $\rho$, plasma flow velocity $v_{x}$, magnetic field component $B_{z}$ and width/position of the primary wave is shown in Figure \ref{kin_prim_wave}. In the density plot (Figure \ref{kin_prim_wave}a) we observe that the wave amplitude remains more or less constant at a value of about $\rho=1.5$ until $t\approx0.05$. Subsequently, the amplitude decreases approximately linearly, as expected \citep{Vrsnak_Lulic2000}, to a value of about $\rho=1.4$ at $t=0.2$, where the primary wave starts entering the CH. A similar decrease is evident in Figure \ref{kin_prim_wave}d and Figure \ref{kin_prim_wave}f for plasma flow velocity and magnetic field component in $z$-direction, where we find a decrease from $v_{x}\approx0.45$ and $B_{z}\approx1.5$ at $t=0$ to $v_{x}\approx0.35$ and $B_{z}\approx1.4$ at $t=0.2$. By looking at the evolution of the width of the wave (see Figure \ref{kin_prim_wave}c), we can see that it remains constant until about $t=0.05$, followed by a slowly increasing width, \ie\ $width_{wave}\approx0.08$ ($t=0$) increases to $width_{wave}\approx0.13$ ($t=0.2$). Similar to Figure \ref{kinem_no_CH}, the decrease of the density amplitude and broadening of the wave coincide with a decrease of plasma flow velocity $v_{x}$ and magnetic field $B_{z}$. In Figure \ref{kin_prim_wave}b we show the position of the amplitude versus time. For the phase velocity of the wave we observe a decrease from $v_{w}\approx1.75$ ($t\approx0.01$) down to $v_{w}\approx1.4$ ($t\approx0.2$) (see Figure\ref{kin_prim_wave}e). 

\subsubsection{Secondary Waves}

\begin{figure}[ht!]
\centering \includegraphics[width=0.5\textwidth]{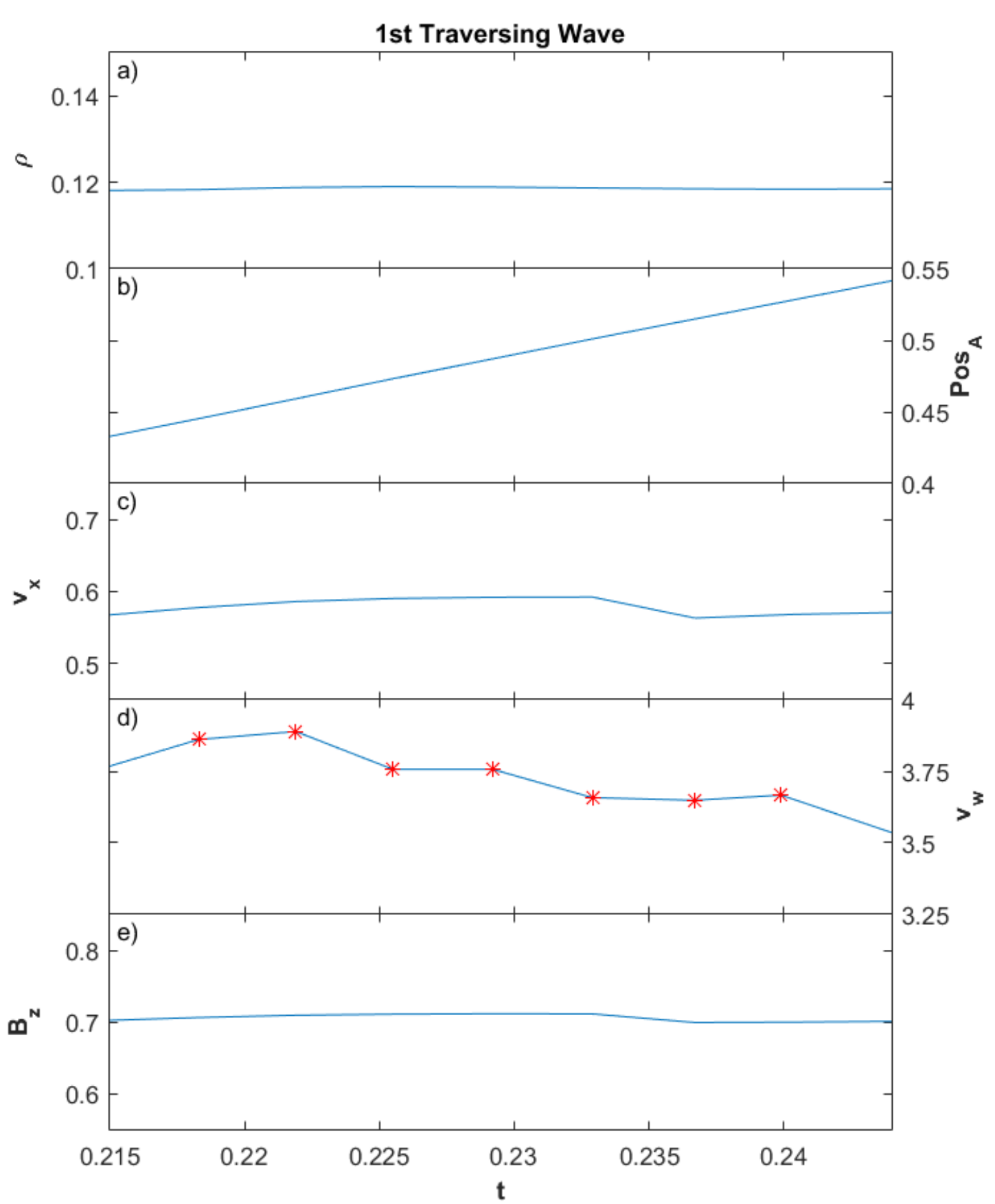}
\caption{From top to bottom: Temporal evolution of density, position of the amplitude, plasma flow velocity, phase velocity and magnetic field of the traversing wave, starting at about $t=0.2$, when the wave is entering the CH and ending at the timestep $t=0.25$, when the wave leaves the CH again.}
\label{kin_travers}
\end{figure}

How the wave components behave while the wave is traversing the CH is shown in Figure \ref{kin_travers}. After entering the CH at $t\approx0.21$ the amplitude quickly decreases from about $\rho\approx1.4$ (see Figure \ref{kin_prim_wave}a at $t=0.2$) to $\rho\approx0.1185$ (see Figure \ref{kin_travers}a), due to the reduced density inside the CH, which corresponds to a decrease from $\Delta\rho=40\%$ to about $\Delta\rho=12\%$. Inside the CH the density amplitude stays approximately constant at $\rho\approx0.12$ (see Figure \ref{kin_travers}a). In Figure \ref{kin_travers}e we observe that the magnetic field component $B_{z}$ also keeps an approximately constant value of $B_{z}\approx0.7$, after having dropped from $B_{y}\approx1.4$ at $t=0.2$ (see Figure\ref{kin_prim_wave}f). By comparing the velocity plots of the primary wave and the traversing wave, it is evident that we observe an increase of phase speed. We find that the phase speed of the primary wave shortly before it enters the CH has a value of $v_{w}\approx1.4$ (see Figure \ref{kin_prim_wave}e at $t\approx0.2$). After the wave has entered the CH, this value increases to $v_{w}\approx3.85$ at $t\approx0.222$ and subsequently decreases slightly to $v_{w}\approx3.55$ at $t\approx0.25$ where one part of the traversing wave leaves the CH again (see Figure \ref{kin_travers}d). In real numbers, by assuming that the Alfv\'{e}n speed is $v_{A}=300$ km s$^{-1}$, this corresponds to an increase of the phase speed from $v_{w}\approx450$ km s$^{-1}$ (shortly before entering the CH) up to $v_{w}\approx1140$ km~s$^{-1}$ inside the CH. There will be no detailed kinematic analysis of the second and third traversing wave  that are reflected inside the CH at the CH boundaries (morphology description in Section 3.2) because it is difficult to detect the amplitude values of the different parameters inside the CH (see Figure \ref{morphology_IN_CH_no2} and Figure \ref{morphology_IN_CH_no3}).

\begin{figure}[ht!]
\centering \includegraphics[width=0.5\textwidth]{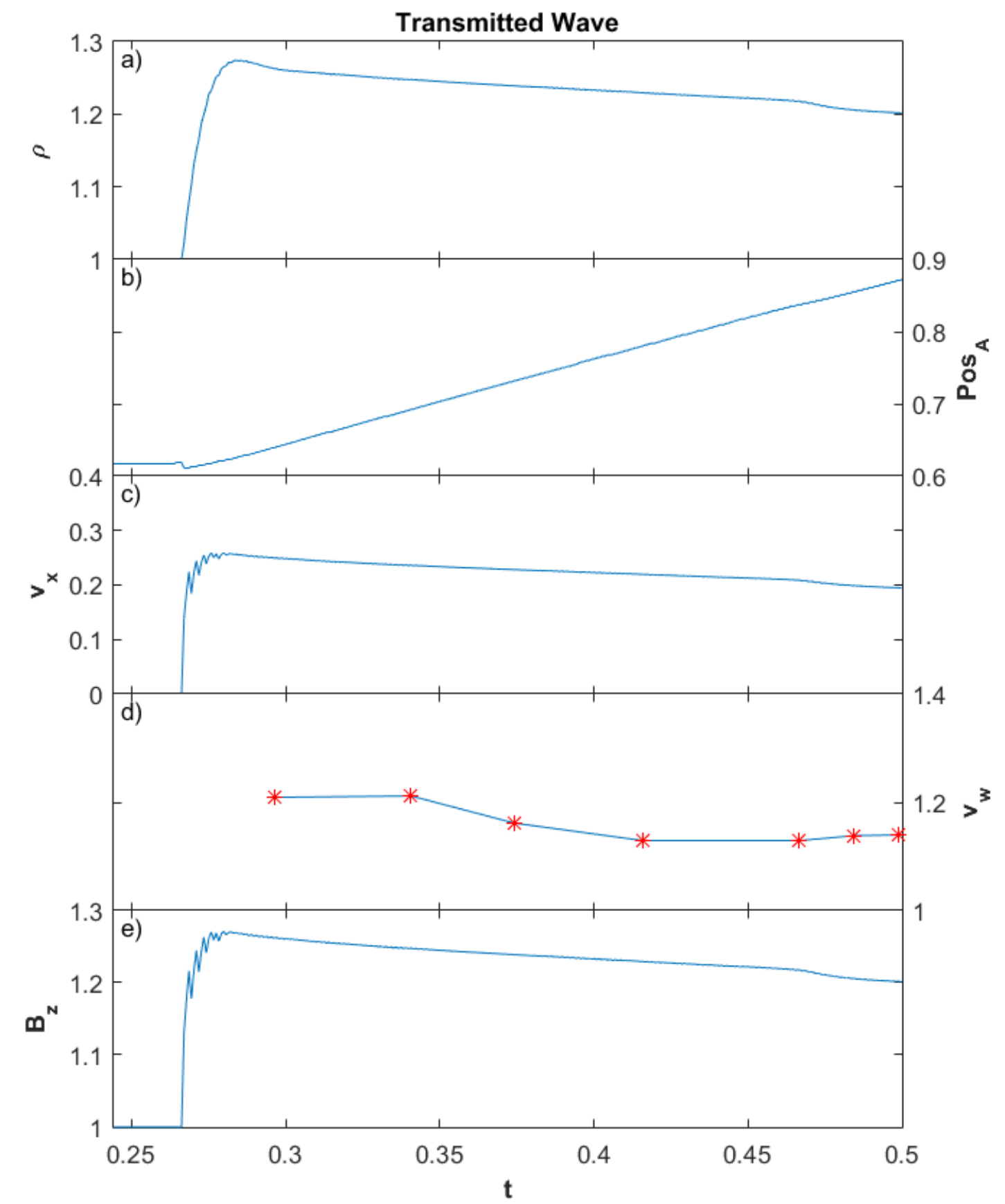} 
\caption{From top to bottom: Temporal evolution of density, position of the amplitude, plasma flow velocity, phase velocity and magnetic field of the transmitted wave, beginning shortly before the wave leaves the CH at about $t=0.255$ and lasting until the end of the simulation run at $t=0.5$.}
\label{kin_transm}
\end{figure}

The kinematics of the transmitted wave are described in Figure \ref{kin_transm}. At about  $t\approx0.28$ the wave leaves the CH and the density amplitude reaches a value of $\rho\approx1.26$ (see Figure  \ref{kin_transm}a). Subsequently, the amplitude decreases to a value of approximately $\rho\approx1.2$, while it is moving further in positive $x$-direction (see Figure \ref{kin_transm}b). As the density's amplitude decreases, also phase speed, plasma velocity and magnetic field component in $z$-direction decrease from $v_{x}\approx0.25$, $v_{w}\approx1.2$ and $B_{z}\approx1.15$ at $t\approx0.28$ to  $v_{x}\approx0.2$, $v_{w}\approx1.15$ and $B_{z}\approx1.275$ at $t=0.5$ (see Figures \ref{kin_transm}c, \ref{kin_transm}d and \ref{kin_transm}e). Again, when we consider real numbers and assume an Alfv\'{e}n speed of $v_{A}=300$ km s$^{-1}$, we observe that the phase speed of $v_{w}\approx450$ km s$^{-1}$, shortly before the wave is entering the CH decreases to a phase speed of the transmitted wave $v_{w}\approx360$ km s$^{-1}$  immediately after the wave has left the CH again, and finally decreases to a value of about $v_{w}\approx345$ km~s$^{-1}$ (which corresponds to  $v_{w}\approx1.15$ in Figure \ref{kin_transm}d) at the end of the run at $t=0.5$.

\begin{figure}[ht!]
\centering \includegraphics[width=0.5\textwidth]{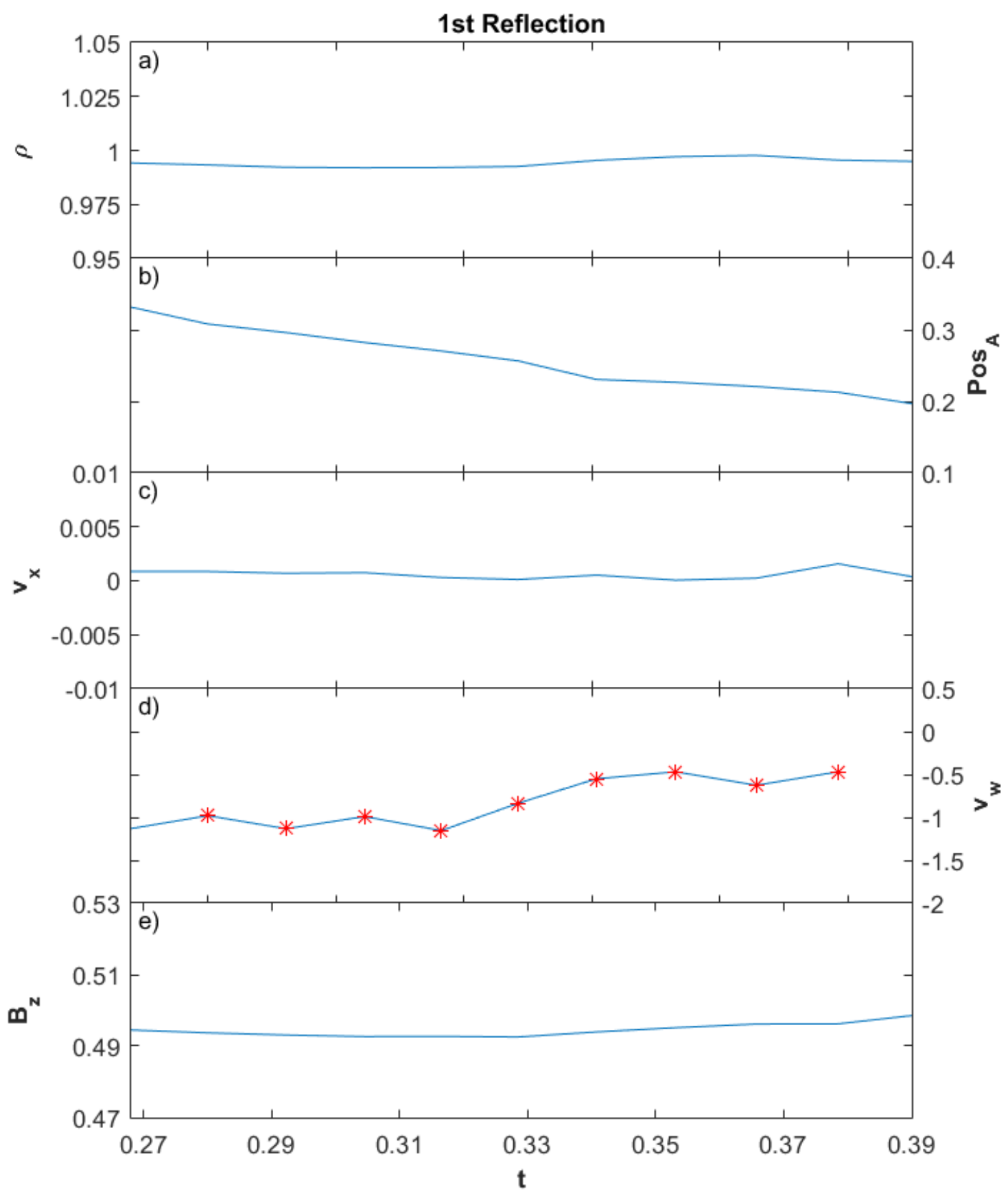}
\caption{From top to bottom: Temporal evolution of density, position of the amplitude, plasma flow velocity, phase velocity and magnetic field of the first reflection starting at ($t\approx0.27$) and ending at ($t=0.39$). }
\label{refl_immed}
\end{figure}

\begin{figure}[ht!]
\centering \includegraphics[width=0.5\textwidth]{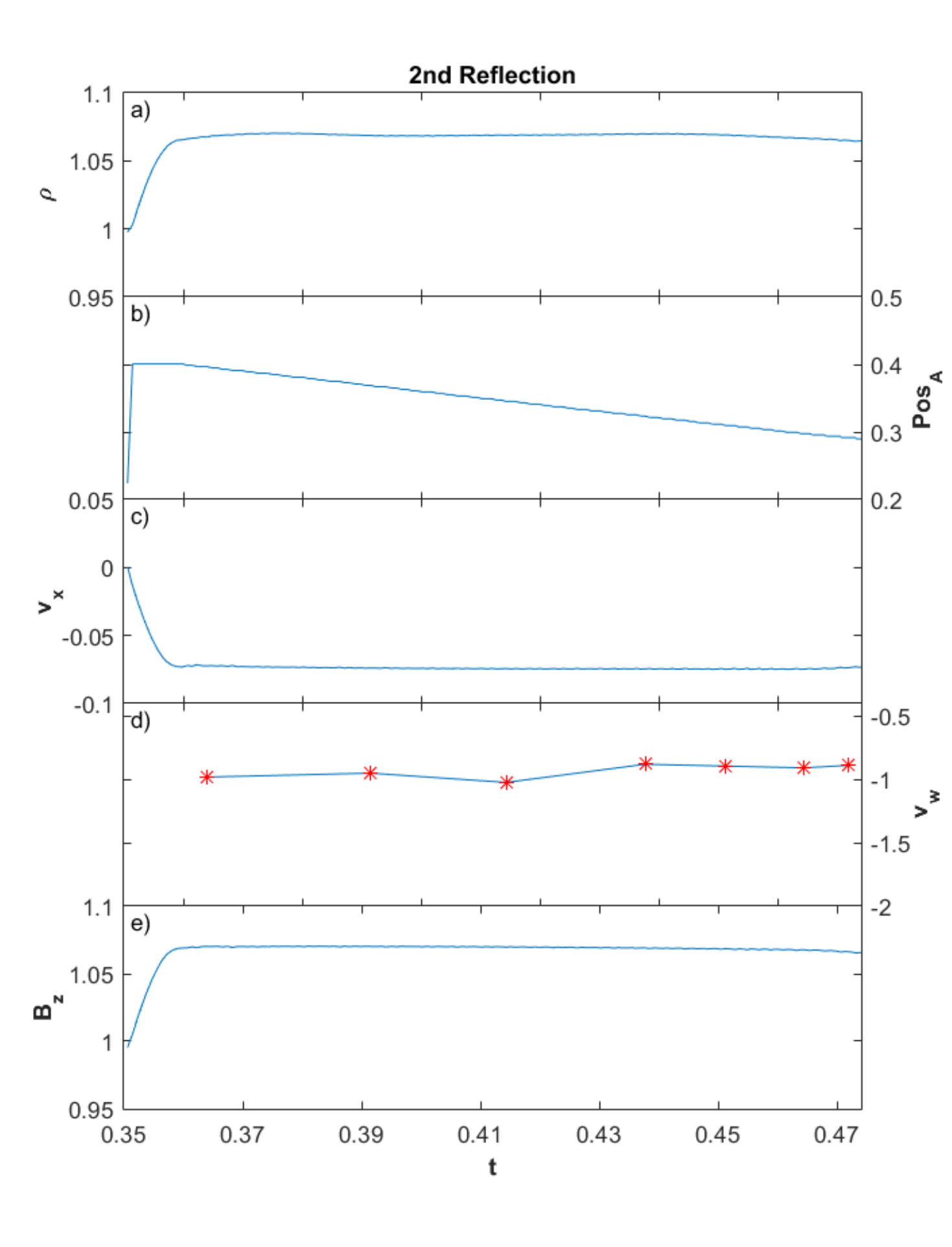}
\caption{From top to bottom: Temporal evolution of density, position of the amplitude, plasma flow velocity, phase velocity and magnetic field of the second reflection, starting shortly before the reflection occurs at ($t\approx0.35$) and ending at the end of the simulation run ($t=0.5$). }
\label{kin_refl_3}
\end{figure}

As already described in Section 3.2, the first reflection like feature is an immediate response of the primary wave's impact on the CH boundary. Due to superposition this feature is kept on the same position ($x\approx0.4$) for some time, until its density amplitude decreases and the reflection is able to move further in negative $x$-direction, ahead of the density depletion. The stationary part of this reflection will be analyzed in Section 4.2.3, whereas the kinematic analysis of the moving part will be done in this Section. We start analyzing this first moving reflection at $t\approx0.27$, the time when we start observing a propagation in negative $x$-direction. In Figure \ref{refl_immed}a we see that the density amplitude remains approximately constant at $\rho\approx1.0$ until $t\approx0.39$. At that time the reflection reaches the area of oscillations caused by numerical effects (see Section 4.1) and we are not able to detect the amplitude values anymore. Figure \ref{refl_immed}b shows how the reflection is moving in negative $x$-direction. In Figure \ref{refl_immed}c and Figure \ref{refl_immed}e we find that plasma flow velocity and magnetic field component in $z$-direction stay approximately constant at values $v_{x}\approx0.001$ and $B_{z}\approx0.495$. The phase speed decreases from $v_{w}\approx-1.1$ to $v_{w}\approx-0.5$ which corresponds to a decrease from $v_{w}\approx-330$ km s$^{-1}$ to $v_{w}\approx-150$ km s$^{-1}$, assuming that we have an Alfv\'{e}n speed of $300$ km s$^{-1}$.

A kinematic analysis of the second reflection, caused by the traversing wave leaving the CH at the left side, is performed in Figure \ref{kin_refl_3}. As we can see in Figure \ref{kin_refl_3}a, \ref{kin_refl_3}c and \ref{kin_refl_3}f, the peak values of density, plasma flow velocity and magnetic field component in $z$-direction are approximately constant, \ie\ $\rho\approx1.07$, $v_{x}\approx-0.07$ and $B_{z}\approx1.07$ from the time step where the reflection occurs ($t\approx0.0355$) until the end of the simulation run. In Figure \ref{kin_refl_3}b it is evident that the reflected wave is moving in negative $x$-direction. The phase speed decreases gradually from $v_{w}\approx1.1$ (at $t\approx0.0355$) to about $v_{w}\approx0.6$ (at $t\approx0.47$) (see Figure \ref{kin_refl_3}d). Assuming an Alfv\'{e}n speed of  $v_{A}=300$ km s$^{-1}$, this corresponds to a decrease of the reflected wave's phase speed from $v_{w}\approx330$ km s$^{-1}$ to $v_{w}\approx180$ km s$^{-1}$. By comparing the phase speed of the primary wave shortly before the wave has entered the CH, $v_{w}\approx450$ km s$^{-1}$ (see Figure \ref{kin_prim_wave}e) and the initial phase speed of the second reflection  $v_{w}\approx-300$ km s$^{-1}$ (see Figure \ref{kin_refl_3}d) we find a decrease of the phase speed of about $33\%$.
In Table 1 we compare the initial $v_{w}(I)$ and ending phase velocities $v_{w}(E)$ of all secondary waves. We find that both reflections and transmitted waves have similar initial speed, which is much slower than the ending phase speed of the primary wave. The initial speed of the first traversing wave inside the CH is about double the speed of the ending speed of the primary wave. Table 2 shows a comparison of the mean phase velocities of the different traversing waves inside the CH. After each reflection at the CH boundary one can observe a change of propagation direction and a decrease of the mean phase velocity of the traversing waves.

\begin{figure*}[ht!]
\centering \includegraphics[width=0.87\textwidth]{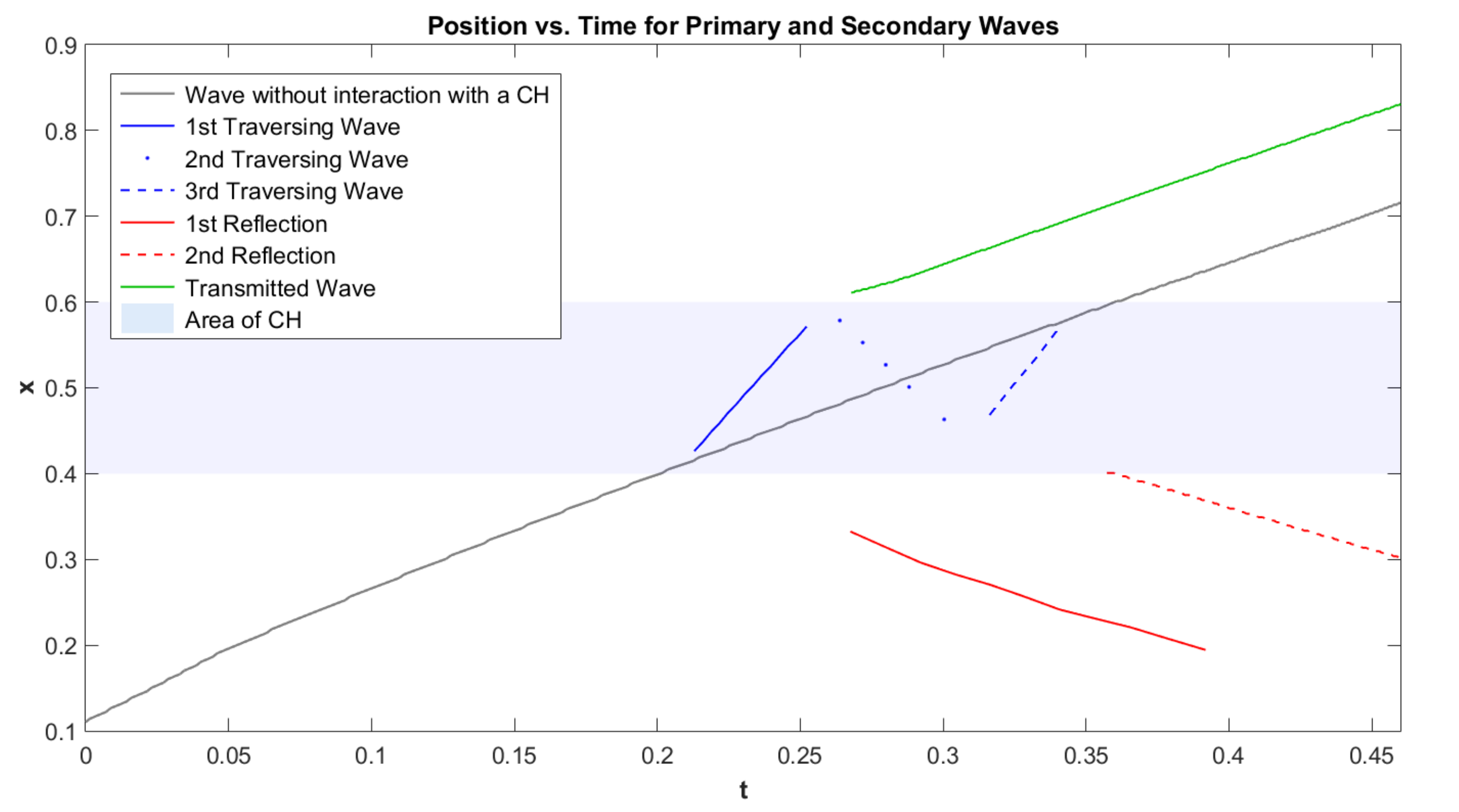}
\caption{Position versus time plot for primary and secondary waves. The gray solid line shows the wave propagation without any interaction with a CH. The light blue area describes the CH. The blue lines denote the different traversing waves: First Traversing Wave (blue solid line), Second Traversing Wave (blue dotted line) and Third Traversing Wave (blue dashed line). The red lines describe reflections of the primary wave at the CH boundary: First Reflection (red solid line), Second Reflection (red dotted line). The green solid line represents the transmitted wave.}
\label{pos_versus_time}
\end{figure*}

Figure \ref{pos_versus_time} shows the temporal evolution of the amplitude position of primary and secondary waves as well as the amplitude position of the wave without any interaction with a CH. The gray solid line describes, on the one hand, the amplitude position of the wave without any interaction with a CH, and on the other hand, it also desribes the position of the primary wave until $t\approx0.2$, which is the time where the entry phase of the primary wave into the CH starts. The blue solid, dotted and dashed lines denote the different traversing waves that are also reflected at the CH boundary inside the CH. The red solid and dashed lines describe the reflections, whereas the green solid line describes the amplitude position of the transmitted wave in time. Shortly after $t=0.2$ we can see that the first traversing wave (blue solid line) is propagating through the CH, subsequently getting reflected at the right CH boundary inside the CH and finally propagating in negative $x$-direction through the CH (blue dotted line). This second traversing wave also gets reflected, but this time at the left CH boundary inside the CH (at $t\approx0.31$) and subsequently propagating in positive $x$-direction (blue dashed line). Moreover, the first reflection (red solid line) starts propagating into negative $x$-direction at about $t=0.27$. It is evident that there is a delay between the time the primary wave (gray solid line) starts entering the CH at $t=0.2$ and the beginning of the first reflection at $t=0.27$. This time shift can be explained by the superposition of the flow associated with the primary wave and the first reflection at the CH boundary, which prevents the first reflection from propagating immediately in negative $x$-direction. The second reflection (red dashed line) starts appearing at about $t=0.35$, as a consequence of the third traversing wave (blue dashed line) partially leaving the CH. At $t=0.27$ the transmitted wave (green solid line) occurs as a result of parts of the first traversing wave (blue solid line) leaving the CH.

\begin{table*}[ht!]
\begin{center}
\begin{tabular*}{0.54\textwidth}{lcccc}
\hline
Feature type &\multicolumn{2}{c}{dimensionless values} &\multicolumn{2}{c}{real numbers $[km s^{-1}]$} \\
  & $v_{w}(I)\approx$ & $v_{w}(E)\approx$ & $v_{w}(I)\approx$ & $v_{w}(E)\approx$ \\
\hline
Primary Wave & 1.75 & 1.5 & 525 & 450 \\
1st Reflection & -1.13 & -0.5 & -339 & -150 \\
2nd Reflection & -1.0 & -0.8 & -300 & -240 \\
1st Traversing Wave & 3.8 & 3.5 & 1140 & 1125 \\
Transmitted Wave & 1.2 & 1.15 & 360 & 345 \\
\hline

\end{tabular*}
\end{center}
\caption{Comparison of the initial $v_{w}(I)$ and ending phase velocities $v_{w}(E) $ of primary and different secondary waves. The first reflection at the CH boundary occurs as an immediate result of the impact of the wave on the CH. The second reflectionn appears at a later time as a consequence of parts of the traversing wave leaving the CH. The first traversing wave is the wave that is moving inside the CH towards positive $x$-direction, until it is reflected for the first time at the right CH boundary inside the CH.}
\end{table*} 

\begin{table*}
\begin{center}
\begin{tabular*}{0.558\textwidth}{lcc}

\hline
Feature type & dimensionless values & real numbers $[km s^{-1}]$\\
 & $v_{w}\approx$  & $v_{w}\approx$ \\
\hline
1st Traversing Wave & 3.596  & 1078\\
2nd Traversing Wave  & -3.097  & -929\\
3rd Traversing Wave & 2.680  & 804\\
\hline

\end{tabular*}
\end{center}
\caption{Comparison of the mean phase velocities of the different traversing waves inside the CH. The first traversing wave starts propagating inside the CH after the primary wave has started entered the CH and is moving further in positive $x$-direction. Second and third traversing waves appear as a consequence of the reflection at the  right and left CH boundary inside the CH.}
\end{table*} 

\begin{figure}[ht!]
\centering \includegraphics[width=0.5\textwidth]{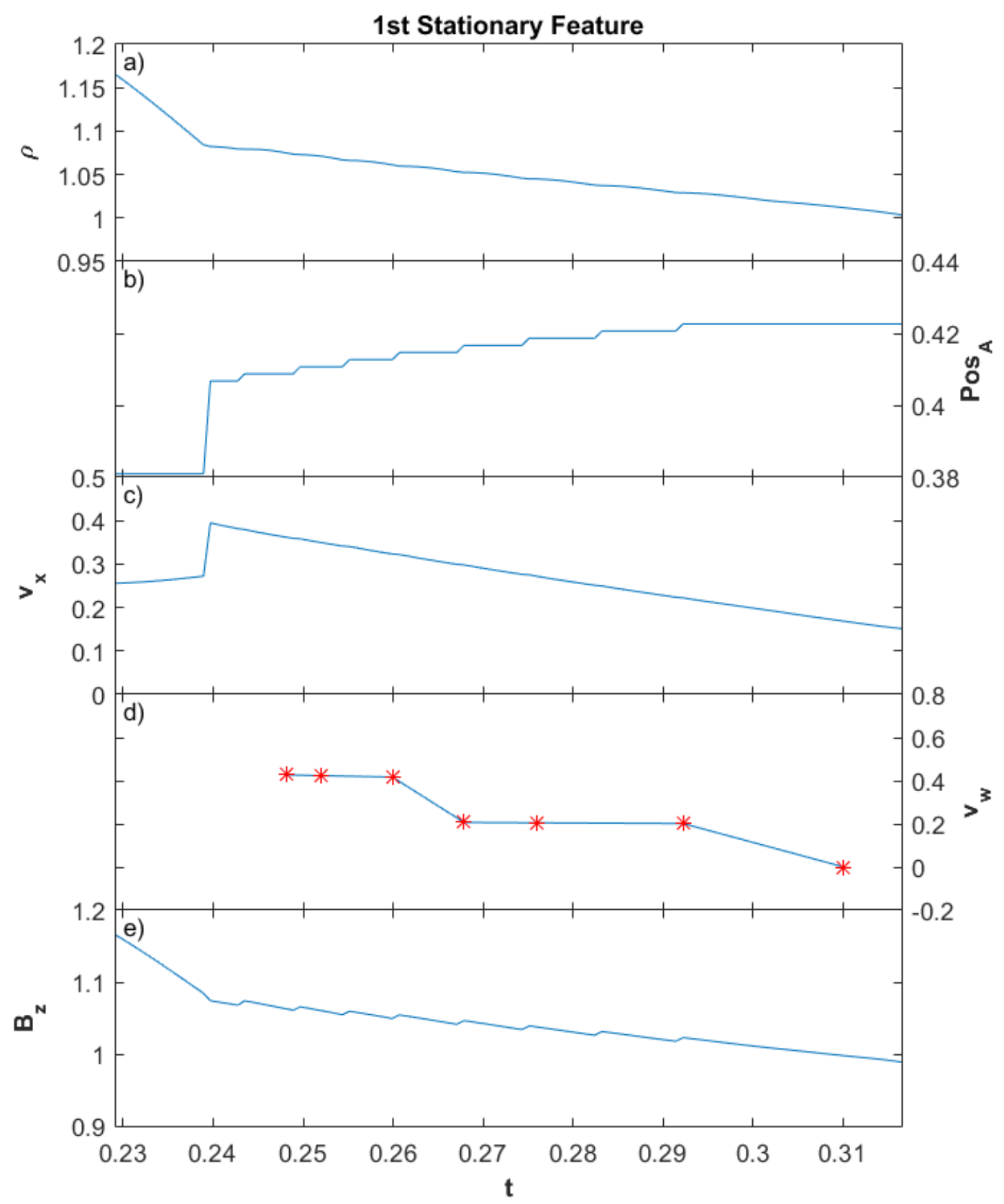}
\caption{From top to bottom: Temporal evolution of density, position of the amplitude, plasma flow velocity, phase velocity and magnetic field of the first stationary feature at the left hand side of the CH.}
\label{kin_refl_1}
\end{figure}

\begin{figure}[ht!]
\centering \includegraphics[width=0.5\textwidth]{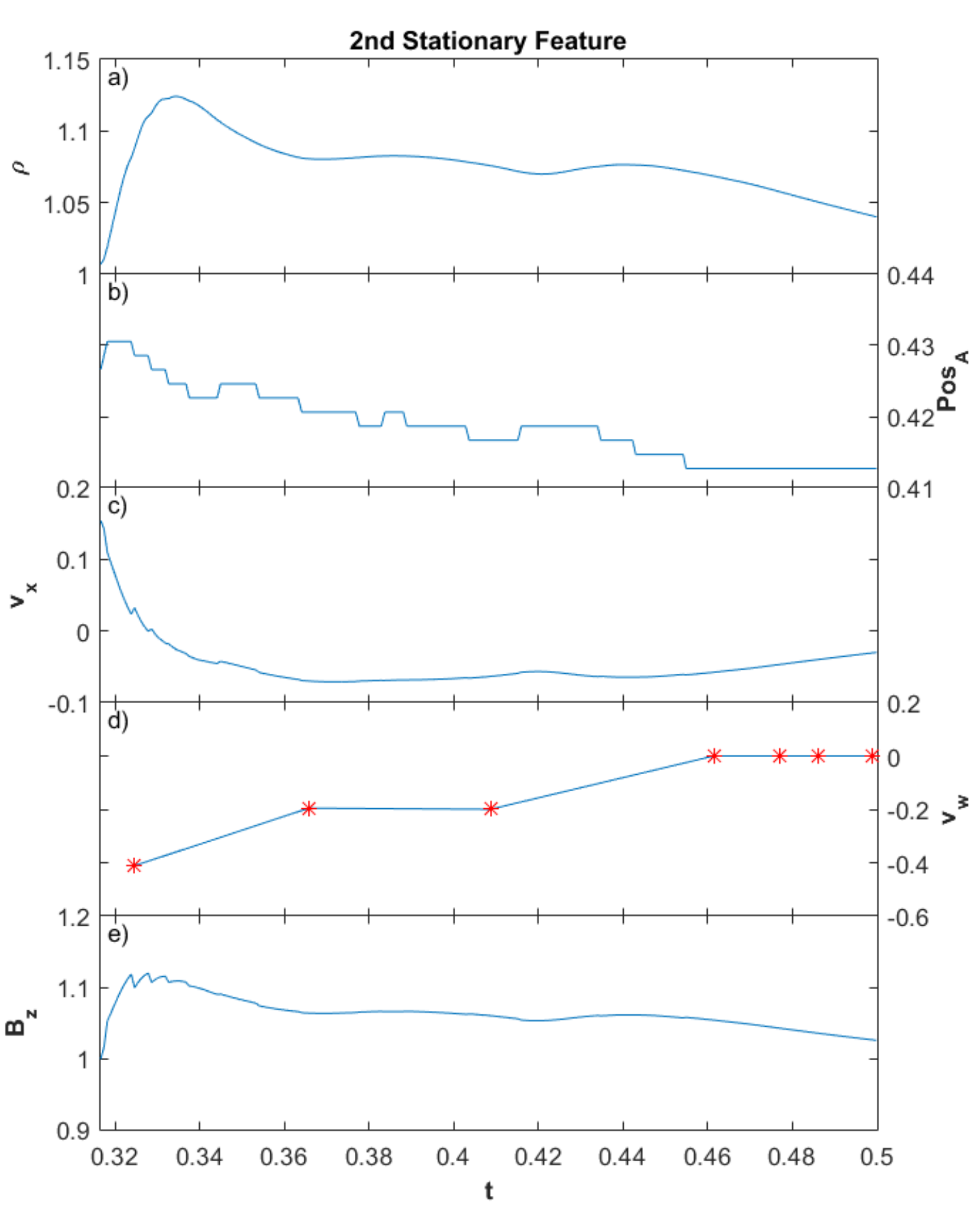}
\caption{From top to bottom: Temporal evolution of density, position of the amplitude, plasma flow velocity, phase velocity and magnetic field of the second stationary feature at the left hand side of the CH.}
\label{kin_refl_2}
\end{figure}

\subsubsection{Stationary Features}

The kinematics of the first stationary feature are described in Figure \ref{kin_refl_1}.
Shortly before $t=0.24$, this feature occurs at the left side of the CH and is subsequently moving slighty into positive $x$-direction (see Figure \ref{kin_refl_1}b). The density plot in Figure \ref{kin_refl_1}a shows a decrease of the amplitude from $\rho\approx1.1$ (at $t\approx0.24$) down to a value of approximately $\rho\approx1$ (at $t\approx0.32$). In the same time interval we also find a small decrease of plasma flow velocity, phase speed and magnetic field component in $z$-direction, \ie\ the values $v_{x}\approx0.4$, $v_{w}\approx1.7$ and $B_{z}\approx1.07$ (at $t\approx0.24$) decrease to $v_{x}\approx0.15$, $v_{w}\approx1.35$ and $B_{z}\approx1.0$ (at $t\approx0.32$) (see Figures \ref{kin_refl_1}c, \ref{kin_refl_1}d and \ref{kin_refl_1}e).

Figure \ref{kin_refl_2} represents the kinematics of the second stationary feature which
appears as another stationary peak at the left CH boundary. This peak starts rising
quickly at about $t\approx0.32$. In contrast to the first stationary feature it moves slightly in negative $x$-direction (see Figure \ref{kin_refl_2}b). Figure \ref{kin_refl_2}a shows how the peak value of the density is decreasing from $\rho\approx1.125$ at $t\approx0.333$ to $\rho\approx1.07$ at $t\approx0.44$, where it reaches the value of the approximately constant density of the second reflected wave (compare to Figure \ref{kin_refl_3}a and Figure\ref{morphology_2} at $t=0.5$). A similar decreasing behaviour to that found for the density, can be seen for the magnetic field component in $z$-direction, where the peak value decreases from $B_{z}\approx1.1$ at $t\approx0.333$ to $B_{z}\approx1.05$ at $t\approx0.44$ (see Figure \ref{kin_refl_2}e). The oscillations in the magnetic field component in the time range $0.32<t<0.34$ can be explained by a certain threshold of the parameter tracking algorithm. Figures \ref{kin_refl_2}c and \ref{kin_refl_2}d show the temporal evolution of plasma flow velocity and phase speed. One can observe an approximately constant plasma flow velocity of $v_{x}\approx-0.05$ and a decrease of the phase speed from $v_{w}=-0.35$ at $t=0.335$ to $v_{w}=0$ at $t=0.5$.

\subsubsection{Density Depletion}

\begin{figure}
\centering \includegraphics[width=0.5\textwidth]{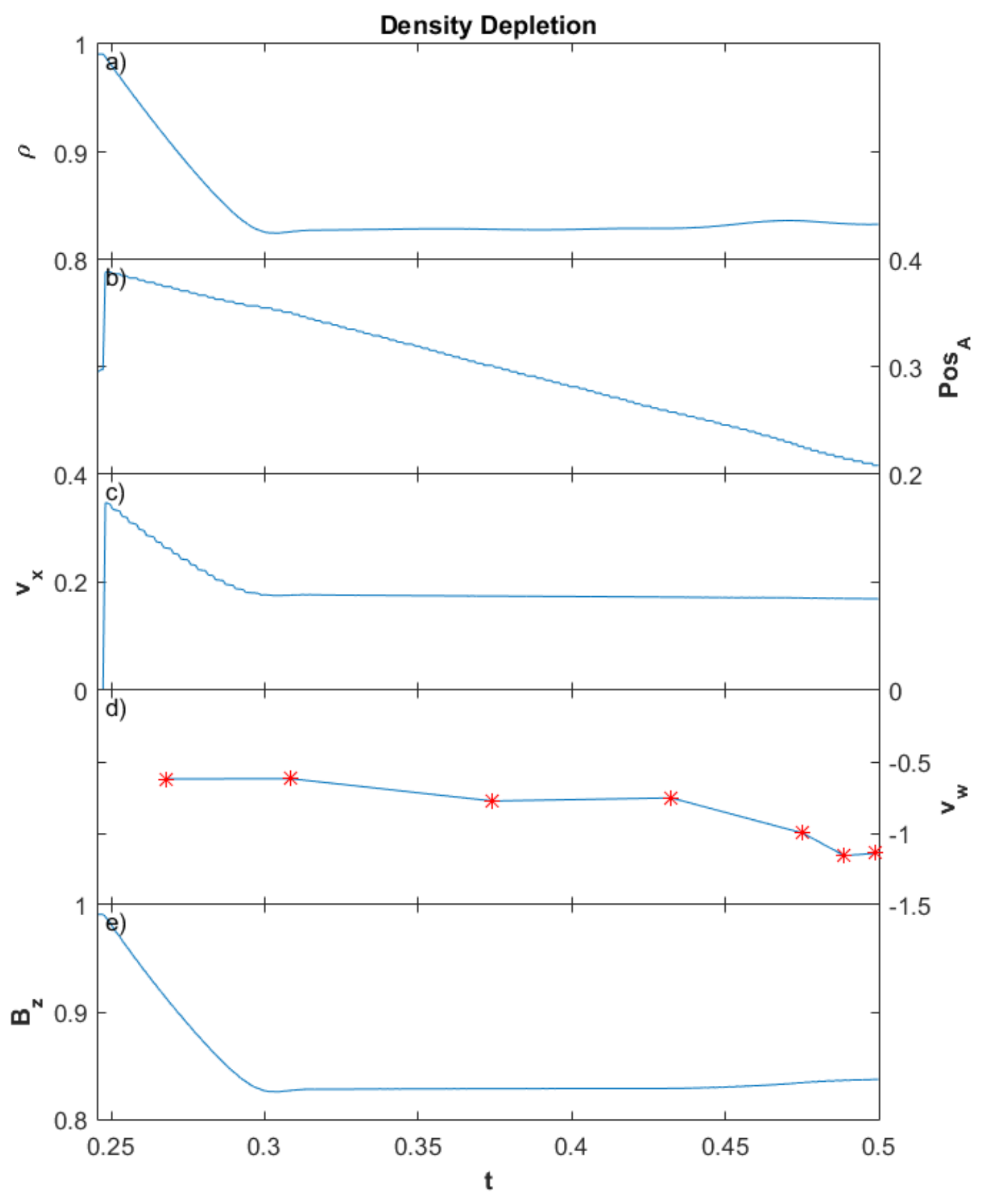} 
\caption{From top to bottom: Temporal evolution of density, position of the amplitude, plasma flow velocity, phase velocity and magnetic field of the moving density depletion.}
\label{kin_min}
\end{figure}

The temporal behaviour of the density depletion is
analyzed in Figure \ref{kin_min}. This feature appears at about $t\approx0.25$, having a
density of about $\rho=0.99$, decreases to a density value of approximately $\rho\approx0.83$ 
until $t\approx0.3$ and then remains at that value until the end of the run (see Figure \ref{kin_min}a). Similar to the density evolution, we find a decrease of the magnetic field component in $z$-direction, from $B_{z}\approx1.0$ at $t\approx0.25$ to $B_{z}\approx0.83$ for $0.3<t<0.5$ (see Figure \ref{kin_min}e). At the same time, while the density is decreasing, the depletion is moving in negative $x$-direction (see Figure \ref{kin_min}b) with a more or less constant plasma flow velocity of $v_{x}\approx0.19$ (see Figure \ref{kin_min}c) and a phase velocity which decreases from $v_{w}\approx1.2$ to about $v_{w}\approx0.5$ (see Figure \ref{kin_min}d). 

\newpage
\newpage

\section{Discussion and Conclusions}

We present a 2.5D simulation of an MHD wave propagation and its interaction with a CH by using a newly developed MHD code. We show that the impact of the incoming wave on the CH leads to different effects:

\begin{itemize}
\item We find different kinds of secondary waves, such as reflections, transmitted waves and waves that are traversing through the CH. The reflections occur, on the one hand, due to an immediate response of the primary wave's impact on the CH boundary. On the other hand, they are a result of the traversing waves being reflected at the CH boundaries inside the CH.
\item We observe stationary features at the CH boundary, where the primary wave is entering the CH. One of these stationary features is caused by the superposition of the flow associated with the primary wave and the first reflection at the CH boundary. Another one appears at a later time, probably a consequence of the traversing wave partially leaving the CH at that position. These features can be interpreted as stationary brightenings in the observations.
\item We have revealed the existence of a density depletion in our simulation, that is moving inbetween the two reflections in negative $x$-direction.
\item We have demonstrated that the primary wave pushes the left CH boundary in direction of the primary wave propagation, caused by the wave front exerting dynamic pressure on the CH.
\item We compared initial and ending phase speeds of secondary waves and their primary counterparts. Moreover, we showed how the reflection, transmission and the traverse of the wave through the CH influence the phase speeds.
\end{itemize}

The findings in our simulation strongly support the wave interpretation of large-scale propagating disturbances in the corona, since, on the one hand, effects like reflection or transmission can be explained by a wave theory but not by a pseudo-wave approach. On the other hand, the simulation indicates that the interaction of an MHD wave with a CH is able to trigger stationary features, which were originally one of the primary reasons for the development of a pseudo-wave theory.

By comparing the phase velocities of the different secondary waves, we find that the initial phase speed of the traversing wave is about double the phase speed of the final phase speed of the primary wave. The initial phase velocities of the reflections and the transmitted wave are of similar size, but about $20 - 30 \%$ smaller than the ending velocity of the primary wave. What we have to be aware of is the fact that the reflections at the CH boundary interact with segments in the rear of the primary wave that are still moving toward the CH. This is a reason for a slow phase speed of the reflected waves in simulations, as well as in observations, where we observe a small difference between the ending velocity of the primary wave and the intial velocity of the reflected wave \citep{Kienreich_etal2012}. 

The density evolution before, during and after the wave penetration of the CH shows that the amplitude of the primary wave decreases about $15\%$ while moving toward the CH. Subsequently, when the wave enters the CH, the density amplitude drops to approximately $8\%$ of the density amplitude value shortly before the wave penetration. After having left the CH, the density amplitude increases up to $82\%$ of the final density amplitude ot the primary wave. The small density amplitude inside the CH illustrates why it is difficult to detect traversing waves in the observations.

Additionally, we showed that the primary wave is capable of pushing the left CH boundary into the direction of wave propagation. Appropriate comparisons between the simulation and observations seem to be difficult in the case of position tracking of CH boundaries, since at present, defining an exact CH boundary in the observations is still a difficult task, especially due to projection effects.

Apart from all these findings, we have to keep in mind that we are describing an idealized situation with many constraints, such as a homogeneous magnetic field, the fact that the pressure is equal to zero over the whole computational box, the assumption of a certain value for the initial wave amplitude and a simplified shape of the CH, where the wave is approaching exactly perpendicular to the CH boundary in every point.

\acknowledgments

The authors gratefully acknowledge the helpful comments from the anonymous referee that have very much improved the quality of this paper. This work was supported by the Austrian Science Fund (FWF): P23618 and P27765. B.V. acknowledges financial support by Croatian Science Foundation under the project 6212 „Solar and Stellar Variability“. I.P. is grateful to the \"OAD (\"Osterreichischer Austauschdienst) and the MŠMT (Ministry of Education, Youth and Sports, Czech Republic) for financing research stays at the Astronomical Institute of the Czech Academy of Sciences in Ond\v{r}ejov.  I.P. acknowledges Ewan C. Dickson for the proof read of this manuscript and the support received by the CREEPY YOUNGSTERS GROUP from the Physics department at the University of Graz. The authors gratefully acknowledge support from NAWI Graz. 

\bibliography{references}

\end{document}